\documentclass[twocolumn]{aastex7}
\newcommand{\edit}[1]{#1}

\newcommand{\ICO}{{I_{\rm CO}}}

\newcommand{\Icont}{{I_{\rm F200W}}}
\newcommand{\Iblueraw}{{I_{\rm F770W}}}
\newcommand{\Iblue}{{I_{\rm F770W, PAH}}}
\newcommand{\Igreen}{{I_{\rm F1130W}}}
\newcommand{\Ired}{{I_{\rm F2100W}}}

\newcommand{\blueband}{{\rm F770W}_{\rm PAH}}
\newcommand{\greenband}{{\rm F1130W}}
\newcommand{\redband}{{\rm F2100W}}

\newcommand{\betablue}{\beta_{\blueband}}
\newcommand{\betagreen}{\beta_{\greenband}}
\newcommand{\betared}{\beta_{\redband}}
\newcommand{\betaCO}{\beta_{CO}}

\newcommand{\nPAH}{n_{\rm PAH}}
\newcommand{\ndust}{n_{\rm dust}}
\newcommand{\ngas}{n_{\rm gas, cold}}

\newcommand{\kKS}{k_{\rm KS}}
\newcommand{\bKS}{b_{\rm KS}}
\newcommand{\sigKS}{\sigma_{\rm KS}}

\newcommand{\Hmolecular}{H_2}
\newcommand{\HII}{H~{\sc ii}}
\newcommand{\HI}{H~{\sc i}}
\newcommand{\Halpha}{H$\alpha$~$\lambda$6563}
\newcommand{\Hbeta}{H$\beta$~$\lambda$4861}
\newcommand{\OIII}{[O~{\sc iii}]~$\lambda$5007}
\newcommand{\SII}{[S~{\sc ii}]~$\lambda$$\lambda$6717,6731}
\newcommand{\NII}{[N~{\sc ii}]~$\lambda$6583}

\newcommand{\Mstar}{M_{\star}}
\newcommand{\SFR}{{\rm SFR}}
\newcommand{\sSFR}{{\rm sSFR}}
\newcommand{\LCO}{L_{\rm CO}}
\newcommand{\MHI}{M_{\rm H I}}
\newcommand{\FWHMa}{{\rm FWHM}_{\rm ang.}}
\newcommand{\FWHMp}{{\rm FWHM}_{\rm phy.}}
\newcommand{\lowb}{low-$b$}
\newcommand{\highb}{high-$b$}

\newcommand{\btheta}{\boldsymbol {\theta}}

\newcommand{\bx}{\mathbf{x}}

\newcommand{\err}[2][]{#2_{\text{err}#1}}

\newcommand{\bxerri}{\err[,i]{\bx}}

\newcommand{\yerri}{\err[,i]{y}}
\newcommand{\obs}[2][]{#2_{\text{obs}#1}}

\newcommand{\bxobsi}{\obs[,i]{\bx}}
\newcommand{\yobsi}{\obs[,i]{y}}

\newcommand{\reproject}{{\tt reproject}}
\newcommand{\linmix}{{\tt LINMIX}}
\newcommand{\mlinmix}{{\tt mLINMIX}}
\newcommand{\modr}{{\tt mODR}}
\newcommand{\mlmethod}{{ML based method}}
\newcommand{\ksmethod}{{KS-test based method}}
\newcommand{\raddest}{{\tt raddest}}

\newcommand{\tdks}{2D KS test}
\newcommand{\ctest}{Cram\'er test}

\usepackage{amsmath}
\usepackage{multirow}
\usepackage{enumitem}

\defcitealias{Jing2024}{JL25}

\shorttitle{CO-MIR Correlations}
\shortauthors{Jing \& Li}
\graphicspath{{./}{figures/}}
\begin{document}

\title{Correlations of ALMA CO(2-1) with JWST mid-infrared fluxes down to scale of $\lesssim$100 parsec in nearby star-forming galaxies from PHANGS}

\author[0009-0004-6271-4321]{Tao Jing}
\affiliation{Department of Astronomy, Tsinghua University, Beijing 100084, China}
\email[show]{tao.jing.astro@gmail.com}

\author[0000-0002-8711-8970]{Cheng Li}
\affiliation{Department of Astronomy, Tsinghua University, Beijing 100084, China}
\email[show]{cli2015@tsinghua.edu.cn}

\correspondingauthor{Tao Jing, Cheng Li}

%% Use the \collaboration command to identify collaborations. This command
%% takes an optional argument that is either a number or the word "all"
%% which tells the compiler how many of the authors above the command to
%% show. For example "\collaboration[all]{(DELVE Collaboration)}" wil include
%% all the authors above this command.
%%
%% Mark off the abstract in the ``abstract'' environment. 
\begin{abstract}
We investigate the correlations of CO (2-1) emission ($\ICO$) with PAH ($\Iblue$ and $\Igreen$) and dust ($\Ired$) emission down to scales of $\lesssim$ 100 pc, by applying \raddest, a novel regression technique recently developed by \citet{Jing2024} that effectively handles uncertainties and outliers in datasets, to 19 nearby star-forming galaxies in the PHANGS sample. We find that for the majority of the data points in all galaxies, the scaling of $\ICO$ with $\Iblue$, $\Igreen$, and $\Ired$ can be well described by log-log linear relations, though with substantial dependence on ionization conditions (i.e., \HII-like, composite-like, and AGN-like). Under given ionization conditions, significant galaxy-to-galaxy variations are identified, and are primarily attributed to variations of intercept $b$, which exhibits clear bimodality. \edit{This bimodality is related to the normalized overall host galaxy star formation rate, such as specific star formation and star formation efficiency.} The differences in slope $k$ and intrinsic scatter $\sigma$ across different MIR bands ($\Iblue$, $\Igreen$, and $\Ired$) are minor compared to their galaxy-to-galaxy variations. All parameters ($k$, $b$, and $\sigma$) depend on the spatial scale of measurement, suggesting that the coupling among CO, PAH, and dust is regulated by different mechanisms at varying scales. \edit{We identify deviations from the log–log linear relation in the brightest regions, primarily characterized by a flattening of the slope.} No significant (3$\sigma$) correlations are found between global properties and the best-fit parameters. We discuss the comparison to previous studies and plausible physics behind the statistical results obtained in this work. %We suggest that the obtained results should be compared with simulations incorporating detailed molecular gas, dust, and PAH models to release their full potential on constraining the corresponding physical processes.
\end{abstract}

%% Keywords should appear after the \end{abstract} command. 
%% The AAS Journals now uses Unified Astronomy Thesaurus (UAT) concepts:
%% https://astrothesaurus.org
%% You will be asked to selected these concepts during the submission process
%% but this old "keyword" functionality is maintained in case authors want
%% to include these concepts in their preprints.
%%
%% You can use the \uat command to link your UAT concepts back its source.

%% From the front matter, we move on to the body of the paper.
%% Sections are demarcated by \section and \subsection, respectively.
%% Observe the use of the LaTeX \label
%% command after the \subsection to give a symbolic KEY to the
%% subsection for cross-referencing in a \ref command.
%% You can use LaTeX's \ref and \label commands to keep track of
%% cross-references to sections, equations, tables, and figures.
%% That way, if you change the order of any elements, LaTeX will
%% automatically renumber them.

\section{Introduction} \label{sec:introduction}

A central pursuit in modern galaxy evolution studies is to understand the coevolution of the multiphase interstellar medium (ISM), which is an ecosystem of molecular gas, polycyclic aromatic hydrocarbons (PAHs) and dust, in addition to the ubiquitous warm/hot gas ionized by various mechanisms. Molecular gas traces the dense ISM and directly fuels star formation. PAHs are regarded as the carriers of absorption features, such as the ultraviolet (UV) bump around 2175 \AA\ \citep[e.g.][]{Li2001, Decleir2019, Massa2022, Shivaei2022, Lin2023, Gordon2024, Battisti2025}, and diffuse interstellar bands \citep[e.g.][]{Leger1985, Crawford1985, van1985, Salama1996, Salama2011, Salama2014}. \edit{In addition to their absorption features, PAHs also exhibit vibrational emission features in the mid-infrared (MIR). These emission features originate from the immediate re-radiation of absorbed UV photons, but PAHs themselves are destroyed in environments with strong UV radiation fields. Therefore, they survive and emit efficiently in photodissociation regions, making PAH emission a useful tracer of such regions.} Dust influences the shielding and chemistry of the ISM through its absorption, scattering, and chemical catalytic properties.

The evolution of molecular gas, PAH and dust is interconnected. Stars formed within molecular clouds are the primary source of dust. Dust, in turn, facilitates the shielding of UV photons, thereby promoting the formation of molecular clouds \citep{Stecher1967, Federman1979, Glassgold1985, Krumholz2009, Sternberg2014, Ballesteros-Paredes2020}. Furthermore, the transition from \HI\ to $\Hmolecular{}$ is significantly accelerated on dust surfaces \citep{Federman1979, Krumholz2009, Krumholz2012, Sternberg2014, Ballesteros-Paredes2020}. The aromatization of small dust grains produces PAHs \citep{Jones2013_DustModel}. In addition, the behavior of these components is regulated by ionization conditions, such as radiation from stars, radiation from active galactic nuclei (AGN), and shocks \citep{Voit1992_PAHDestruction,Jones1994_DustDestruction,Zhou2023,Narayanan2023,Guo2025}. Therefore, the relationships among these components encode essential information about the physical state of the ISM, physical and chemical processes that affect ISM, and the regulation of star formation across different environments and scales.

Due to its importance, the correlation of molecular gas with both PAH and dust emission has been extensively studied in recent years, on scales from galaxy sizes down to $\sim$kpc.  For instance, \cite{Jiang2015} and \cite{Gao2019} identify log-log linear scaling relations of molecular gas (traced by CO(1-0) or CO(2-1)\footnote{Unless otherwise specified, CO in this paper refers to $^{12}$CO}) with PAH (traced by the WISE W3 band; \citealt{Wright2010_WISE}) and dust (traced by the WISE W4 band) at galactic scales. They find that the CO(1-0)–WISE W3 band correlation exhibits less scatter compared to the CO(1-0)–WISE W4 correlation. \edit{Spectroscopic studies further confirm these trends by directly probing individual PAH features. \cite{Cortzen2019} report correlations between CO(1-0) emission and the $6.2~\mu$m and $7.7~\mu$m PAH features, while \cite{Whitcomb2023} identify clear CO(2-1)--PAH correlations across multiple PAH features at wavelengths longer than $5~\mu$m using spectroscopy.} \cite{Chown2021} extend the correlation between CO(1-0) and the WISE W3 band to kpc scales and find a slope consistent with that observed at galactic scales. \cite{Leroy2023_1kpc} further validate these results by studying the correlation between CO(1-0) and CO(2-1) with $8~\mu$m (Spitzer; \citealt{Werner2004_Spitzer}), $12~\mu$m (WISE W3), $22~\mu$m (WISE W4), and $24~\mu$m (Spitzer) bands across kpc to galactic scales. Comparing different bands, they report steeper CO vs. mid-infrared slopes for the $8~\mu$m and $12~\mu$m bands than for the $22~\mu$m and $24~\mu$m bands. Additionally, they identify an anti-correlation between the intercept of these scaling relations and the host galaxies' specific star formation rate (sSFR). \cite{Gao2022} find that the W3 band correlates more tightly with $^{12}$CO than with $^{13}$CO, C$^{18}$O, or other dense gas tracers (e.g., HCN(4-3), HCO$^+$(4-3)). More recently, \cite{Gao2025} report that the CO(1-0)–WISE W3 and CO(2-1)–WISE W3 scaling relations exhibit steeper slopes in early-type galaxies compared to star-forming galaxies.

%However, these studies do not investigate these scaling relations at the natural scales of molecular clouds and photodissociation regions ($\lesssim 100$ pc) due to limited spatial resolution. This limitation primarily arises from PAH and dust emission measurements based on WISE or Spitzer.

The launch of JWST \citep{Gardner2006_JWST} has recently enabled both PAH and dust emission to be measured at sub-arcsecond spatial resolution. \edit{By combining PHANGS-JWST \citep{Lee2023_PHANGS_JWST, Williams2024_PHANGS_JWST}, PHANGS-ALMA \citep{Leroy2021_PHANGS_ALMA_pipeline, Leroy2021_PHANGS_ALMA_survey} and additional JWST and ALMA observations}, \cite{Leroy2023} and \cite{Chown2024,Chown2025_PAHinDwarf} are the first to explore the correlations of CO with PAH and dust at (sub-)cloud scales ($\le$100 pc). However, as noted by \citet[][hereafter \citetalias{Jing2024}]{Jing2024}, \edit{conventional regression approaches used in these studies do not explicitly account for all relevant observational uncertainties and outliers, which may affect the inferred scaling relations and obscure the characterization of their intrinsic scatter.} For instance, as demonstrated in \citetalias{Jing2024}, the previously reported sublinear slope of the CO–$\redband$ scaling relation primarily arises from the limited signal-to-noise ratio (S/N) of $\redband$ and the presence of outliers. Moreover, the effect of ionization conditions on these scaling relations remains unexplored. The ionization condition dependence encodes critical information about the influence of stellar or AGN feedback on the coupling of CO, PAH, and dust in the ISM.

In this work, we analyze 19 galaxies in the PHANGS sample, which are included in the PHANGS-JWST, PHANGS-ALMA, and PHANGS-MUSE \citep{Emsellem2022_PHNAGS_MUSE} samples. The optical emission line measurements provided by PHANGS-MUSE enable us to divide the regions into \HII-like, composite-like, and AGN-like ionization subgroups, facilitating the study of the dependence on ionization conditions. For the regression analysis, we adopt the recently developed method \raddest~(\citetalias{Jing2024}). As demonstrated by \citetalias{Jing2024}, \raddest\ achieves the highest accuracy among all existing widely used techniques, particularly for datasets with limited S/N ratios and outliers. This is validated using both mock datasets and observations of four galaxies included in the PHANGS-JWST and PHANGS-ALMA samples. Consequently, the results obtained with this method minimize the influence of limited S/N and outliers, providing unbiased measurements of scaling relationships, including intrinsic scatter. By utilizing the ability of \raddest\ to model any scaling relation formula, we further investigate potential deviations from the log–log linear relation.

This paper is organized as follows. In \autoref{sec:data_methodology}, we describe the data preparation approach and briefly introduce the regression technique. Statistical results are presented in \autoref{sec:result}.  \autoref{sec:discussion} compares our findings to previous studies and explores the plausible physical mechanisms underlying the scaling relations. Finally, we summarize our key findings in \autoref{sec:summary}.

\section{Data and Methodology}\label{sec:data_methodology}

\subsection{The PHANGS galaxy sample} \label{sec:phangs}

\begin{deluxetable*}{ccccccccccc}
	\tablecolumns{11}
	\tablewidth{0pt}
	\tablecaption{Global properties and highest available resolution after PSF matching for the galaxies analyzed in this work \label{tab:galaxies}}
	\tablehead{
		\colhead{ID \tablenotemark{a}} &
		\colhead{Name} &
		\colhead{$\log \Mstar$} &
		\colhead{$\log \SFR$} &
        \colhead{$\log \LCO$} &
		\colhead{$\log \MHI$} &
        \colhead{$D$} &
        \colhead{$i$ \tablenotemark{b}} &
        \colhead{$\FWHMa$ \tablenotemark{c}} &
        \colhead{$\FWHMp$ \tablenotemark{d}} &
        \colhead{\highb} 
        \\
        \colhead{} &
		\colhead{} &
		\colhead{[$M_\odot$]} &
		\colhead{[$M_\odot\ {\rm yr}^{-1}$]} &
        \colhead{[K km s$^{-1}$ pc$^2$]} &
		\colhead{[$M_\odot$]} &
        \colhead{[Mpc]} &
        \colhead{[deg]} &
        \colhead{[arcsec]} &
        \colhead{[pc]} &
        \colhead{} 
	}
	\startdata
	1 & NGC5068 & 9.41 & -0.56 & 7.26 & 8.82 & 5.20 & 35.70 & 1.04 & 26.22 & No \\
    2 & IC5332 & 9.68 & -0.39 & 7.09 & 9.30 & 9.01 & 26.90 & 0.87 & 38.00 & Yes \\
    3 & NGC1087 & 9.94 & 0.11 & 8.32 & 9.10 & 15.85 & 42.90 & 1.60 & 123.15 & No \\
    4 & NGC1385 & 9.98 & 0.32 & 8.37 & 9.19 & 17.22 & 44.00 & 1.27 & 105.86 & No \\
    5 & NGC2835 & 10.00 & 0.10 & 7.71 & 9.48 & 12.22 & 41.30 & 1.15 & 68.13 & No \\
    6 & NGC7496 & 10.00 & 0.35 & 8.33 & 9.07 & 18.72 & 35.90 & 1.68 & 152.02 & No \\
    7 & NGC0628 & 10.34 & 0.24 & 8.41 & 9.70 & 9.84 & 8.90 & 1.12 & 53.49 & Yes \\
    8 & NGC3351 & 10.37 & 0.12 & 8.13 & 8.93 & 9.96 & 45.10 & 1.46 & 70.66 & Yes \\
    9 & NGC4254 & 10.42 & 0.49 & 8.93 & 9.48 & 13.10 & 34.40 & 1.78 & 113.13 & No \\
    10 & NGC4303 & 10.51 & 0.73 & 9.00 & 9.67 & 16.99 & 23.50 & 1.81 & 149.34 & No \\
    11 & NGC4535 & 10.54 & 0.34 & 8.61 & 9.56 & 15.77 & 44.70 & 1.56 & 119.14 & Yes \\
    12 & NGC1300 & 10.62 & 0.07 & 8.50 & 9.38 & 18.99 & 31.80 & 1.23 & 113.11 & Yes \\
    13 & NGC1512 & 10.72 & 0.11 & 8.26 & 9.88 & 18.83 & 42.50 & 1.25 & 114.11 & Yes \\
    14 & NGC1672 & 10.73 & 0.88 & 9.05 & 10.21 & 19.40 & 42.60 & 1.93 & 181.75 & No \\
    15 & NGC4321 & 10.75 & 0.55 & 9.02 & 9.43 & 15.21 & 38.50 & 1.67 & 122.89 & Yes \\
    16 & NGC1566 & 10.79 & 0.66 & 8.89 & 9.80 & 17.69 & 29.50 & 1.25 & 107.57 & No \\
    17 & NGC3627 & 10.84 & 0.59 & 8.98 & 9.09 & 11.32 & 57.30 & 1.63 & 89.25 & No \\
    18 & NGC1433 & 10.87 & 0.05 & 8.47 & 9.40 & 18.63 & 28.60 & 1.10 & 99.14 & Yes \\
    19 & NGC1365 & 11.00 & 1.24 & 9.49 & 9.94 & 19.57 & 55.40 & 1.38 & 130.81 & Yes
	\enddata
	\tablenotetext{a}{Galaxies are ordered by increasing stellar mass $\Mstar$.}
    \tablenotetext{b}{Inclination angle of the galaxy.}
	\tablenotetext{c}{Angular FWHM after PSF matching.}
    \tablenotetext{d}{Physical FWHM after PSF matching.}
\end{deluxetable*}

We consider 19 galaxies from the Physics at High Angular Resolution in Nearby Galaxies Survey (PHANGS). For each galaxy, PHANGS-ALMA \citep{Leroy2021_PHANGS_ALMA_pipeline, Leroy2021_PHANGS_ALMA_survey} provides the CO(2-1) intensity ($\ICO$) map, while PHANGS-JWST\footnote{All the PHANGS-JWST data used in this paper can be found in MAST: \dataset[https://doi.org/10.17909/ew88-jt15]{https://doi.org/10.17909/ew88-jt15}.} \citep{Lee2023_PHANGS_JWST, Williams2024_PHANGS_JWST} provides maps of PAH emission in the F770W ($\Iblueraw$) and F1130W ($\Igreen$) bands, as well as dust emission in the F2100W band ($\Ired$). Following \citet{Chown2024}, we estimate the stellar continuum contamination in the F770W band to be 12\% of the F200W band intensity ($\Icont$). The continuum-subtracted F770W intensity is therefore given by $\Iblue = \Iblueraw - 0.12 \Icont$. \edit{We note that although PAH features contribute significantly to the emission in $\blueband$ and $\greenband$, which we refer to as ``PAH bands'', dust continuum emission also contributes to these bands \citep{Whitcomb2023, Sutter2024_PAHinDust,Chown2025_PAHinMIR, Donnelly2025_PAHinMIR}. While several dust continuum subtraction methods have been proposed in the literature \citep{Chown2025_PAHinMIR, Donnelly2025_PAHinMIR}, their specific implementations depend on the adopted models (e.g., the decomposition of dust and PAH contributions for calibration), local environments, and spatial resolutions. As this work focuses on the statistical analysis of the correlations of CO with PAH and dust emission, we aim to retain the original observables as much as possible while removing contamination from unrelated components. This approach enables direct comparisons with other studies and provides calibrations that are closer to observations for future forward-modeling studies.}
Additionally, PHANGS-MUSE \citep{Emsellem2022_PHNAGS_MUSE} provides optical emission line fluxes, including \Halpha, \Hbeta, \SII, \NII, and \OIII, which we use to diagnose the ionization source in each spaxel. Specifically, we classify spaxels as \HII-like, composite-like, or AGN-like using the $P1-P2$ diagnostic diagram proposed by \citet{Ji2020}. \edit{It is important to note that this classification is based solely on spaxel-level emission-line ratios. Consequently, some spaxels may be classified as AGN-like even in galaxies without confirmed AGN activity, due to measurement uncertainties or local ionization mechanisms that can produce AGN-like line ratios, such as shocks. To assess the impact of such contamination, we repeated our analysis using only AGN-like spaxels located in galaxies known to host AGNs (NGC1365, NGC1566, NGC1672, NGC4303, and NGC7496; \citealt{Baron2025_PAHvsOptical}). We find that all of our key results remain unchanged.}

For each galaxy, we perform Gaussian convolution to match the Point Spread Functions (PSFs) of the intensity and flux maps across all bands. The corresponding uncertainty maps are generated through error propagation, which accounts for correlations between nearby pixels as described by \citet{Klein2021}. All maps are resampled to a unified World Coordinate System (WCS) with a pixel size set to half the PSF's Full Width at Half Maximum (FWHM), using the \reproject\ package \citep{reproject_0p13p1}. Projection corrections are applied to $\ICO$, $\Iblue$, $\Igreen$, and $\Ired$ to account for the inclination of the galaxy. Unless otherwise specified (e.g., in \autoref{sec:diff_spatial_scale}, where we investigate the effects of spatial scale), all analyses are performed on this PSF-matched dataset.

\autoref{tab:galaxies} lists the global properties of the galaxies taken from \citet{Leroy2021_PHANGS_ALMA_survey}, along with the FWHM of the matched PSF and the corresponding physical resolution. All 19 galaxies are located on the star-forming main sequence, with specific star formation rates (sSFR$\equiv$SFR/$M_\ast$) \edit{well above $10^{-11}$ ${\rm yr}^{-1}$},  and have stellar masses in the range of $\sim3\times 10^{9}$–$10^{11}\ M_\odot$.

\subsection{\raddest} \label{sec:regression}

We model the correlation between $\ICO$ and $\Iblue$, $\Igreen$, or $\Ired$ using a log-log linear relation that incorporates Gaussian intrinsic scatter in logarithmic space. The relation is expressed as:
\begin{equation} \label{eq:loglinear}
\log \ICO = k \log I_X + b + \epsilon,~~\epsilon \sim \mathrm{N}(0, \sigma^2),
\end{equation}
where $I_X$ denotes $\Iblue$, $\Igreen$, or $\Ired$. The parameters to be estimated are the slope $k$, the intercept $b$, and the intrinsic scatter $\sigma$. These parameters are determined by applying \raddest, a novel regression code developed by \citetalias{Jing2024}, to the PHANGS galaxy dataset described in the previous subsection.

As a regression method, \raddest\ infers the posterior distribution of the model parameters (i.e., $\btheta = \{k, b, \sigma\}$ in this case) given the observed data. In our case, the data $D = \{D_i\}$ (where $i=1,2,...m$) consist of a set of spaxels within a galaxy. Each data point $D_i$ includes a dependent variable $\yobsi$ (i.e., $\ICO$ from ~\autoref{eq:loglinear}), an independent variable $\bxobsi$ (i.e., $I_X$ from~\autoref{eq:loglinear}), and their corresponding uncertainties, $\yerri$ and $\bxerri$. Assuming the data points are independent, the likelihood of the entire dataset $P(D|\btheta)$ is the product of the individual likelihoods, $P(D_i|\btheta)$. The likelihood for a single data point is derived as follows:
\begin{equation} \label{eq:likelihood_int}
    \begin{aligned}
    P(D_i|\btheta) = & P(\yobsi, \yerri, \bxobsi, \bxerri|\btheta) \\
    = & \int P(y |\bx, \btheta) P(\yobsi, \bxobsi|\yerri, \bxerri, y, \bx)  \\
    & \times  P(\yerri|y) P(\bxerri|\bx) \  P(\bx)\ dy\ d\bx,
    \end{aligned}
\end{equation}
where $P(y|\bx, \btheta)$ is the underlying model describing the correlation between $\bx$ and $y$ (\autoref{eq:loglinear}); $P(\yobsi, \bxobsi|\yerri, \bxerri, y, \bx)$ is the noise model, which we assume to be Gaussian; $P(\yerri|y)$ and $P(\bxerri|\bx)$ describe how the uncertainties $\yerri$ and $\bxerri$ relate to the intrinsic values $y$ and $\bx$, respectively; $P(\bx)$ represents the intrinsic distribution of the independent variable $\bx$. The integral on the right-hand side of the equation is evaluated using Gauss-Hermite quadrature. 

% \begin{equation}
%     \begin{aligned}
%         & P(\{\bxobsi\}, \{\bxerri\}) = \prod_i P (\bxobsi, \bxerri) 
%         \\ & = \prod_i\int P(\bxobsi|\bx, \bxerri) P(\bxerri, \bx) d\bx
%         \\ & =\prod_i\int P(\bxobsi|\bx, \bxerri) 
%         P(\bxerri|\bx)
%         P(\bx)\ d\bx.
%     \end{aligned}
% \end{equation}

The distributions $P(\bxerri|\bx)$ and $P(\bx)$ are generally not known from the observed data. To address this, \raddest\ employs normalizing flows (NFs; \citet{Dinh2014,Rezende2015}) to model them. In practice, these distributions are estimated by maximizing the observable joint distribution $P(\{\bxobsi\}, \{\bxerri\}) = \prod_i \int P(\bxobsi|\bx, \bxerri) P(\bxerri|\bx) P(\bx)\ d\bx$. The goodness-of-fit for the NF is evaluated using \edit{\ctest{}}\footnote{The implementation is provided by the {\tt cramer} package for {\tt R}: \url{https://doi.org/10.32614/CRAN.package.cramer}.}, which compares the generated pairs $(\obs[,\rm gen]{\bx}, \err[,\rm gen]{\bx})$ with the observed pairs $(\obs{\bx}, \err{\bx})$. The NF hyperparameters are tuned to achieve a $p$-value greater than 0.01. The distributions $P(\yerri|y)$ and $P(y)$ are estimated analogously. The estimated $P(\bxerri|\bx)$, $P(\yerri|y)$, and $P(\bx)$ are then substituted into~\autoref{eq:likelihood_int} to calculate the likelihood for individual data points $P(D_i|\btheta)$ and, consequently, the total likelihood $P(D|\btheta)$. 

The likelihood $P(D|\btheta)$ is used for parameter estimation and posterior sampling (under a uniform prior in this work) via two distinct methods. The first method, referred to hereafter as \mlmethod, involves standard maximum a posteriori (MAP) or maximum likelihood (ML) estimation. A gradient descent algorithm finds the MAP/ML solution, and then \edit{a standard Markov chain Monte Carlo (MCMC) or Hamiltonian Monte Carlo (HMC) algorithm samples the posterior distribution}. The second method, termed \ksmethod, leverages the generative capability of NFs. Here, parameters are estimated by minimizing the distribution distance, \edit{quantified by the approximate $p$-value from the two-dimensional Kolmogorov-Smirnov (2D KS) test \citep{Peacock1983, Fasano1987, Press2002}\footnote{Implemented in \url{https://github.com/astro-jingtao/ndtest}}} between generated $(\obs[,\rm gen]{\bx}, \obs[,\rm gen]{y})$ and observed $(\obs{\bx}, \obs{y})$ data pairs. However, obtaining a rigorous posterior sample with this approach is infeasible. Instead, an approximate posterior sample is constructed by generating data points from the prior distribution and using the approximate $p$-value from the \tdks\ as a weight for each point. \edit{It should be emphasized that the \tdks\ is not employed here as a formal hypothesis test. Rather, its approximate $p$-value is used as a sample-based measure of the distance between the mock data generated under a given parameter set and the observed data. The \tdks\ is adopted because it provides a practical balance between computational efficiency and performance in this application.}

As demonstrated by \citetalias{Jing2024}, tests on both mock data and real PHANGS-ALMA and PHANGS-JWST data show that for large samples ($>1000$ spaxels), both the \ksmethod\ and \mlmethod\ significantly outperform existing widely-used methods, particularly in low S/N regimes. Among them, the \ksmethod\ achieves the highest accuracy and robustness, provided the sample size is sufficiently large ($\geq 1000$ spaxels, a condition met in most of our analyses). For intermediate sample sizes (300–1000 spaxels), the \mlmethod\ delivers optimal performance, while in the low-data regime (below 300 spaxels), it remains competitive with other state-of-the-art methods. A GPU-compatible Python implementation of \raddest\ is publicly available\footnote{\raddest\ package: \url{https://github.com/astro-jingtao/raddest}}. Based on these findings, we employ the \ksmethod\ for the majority of our analysis, resorting to the \mlmethod\ only for specific purposes (e.g., to model the behavior of outliers; see the next subsection).

\begin{figure*}[ht!]
    \centering
    \includegraphics[width=\textwidth]{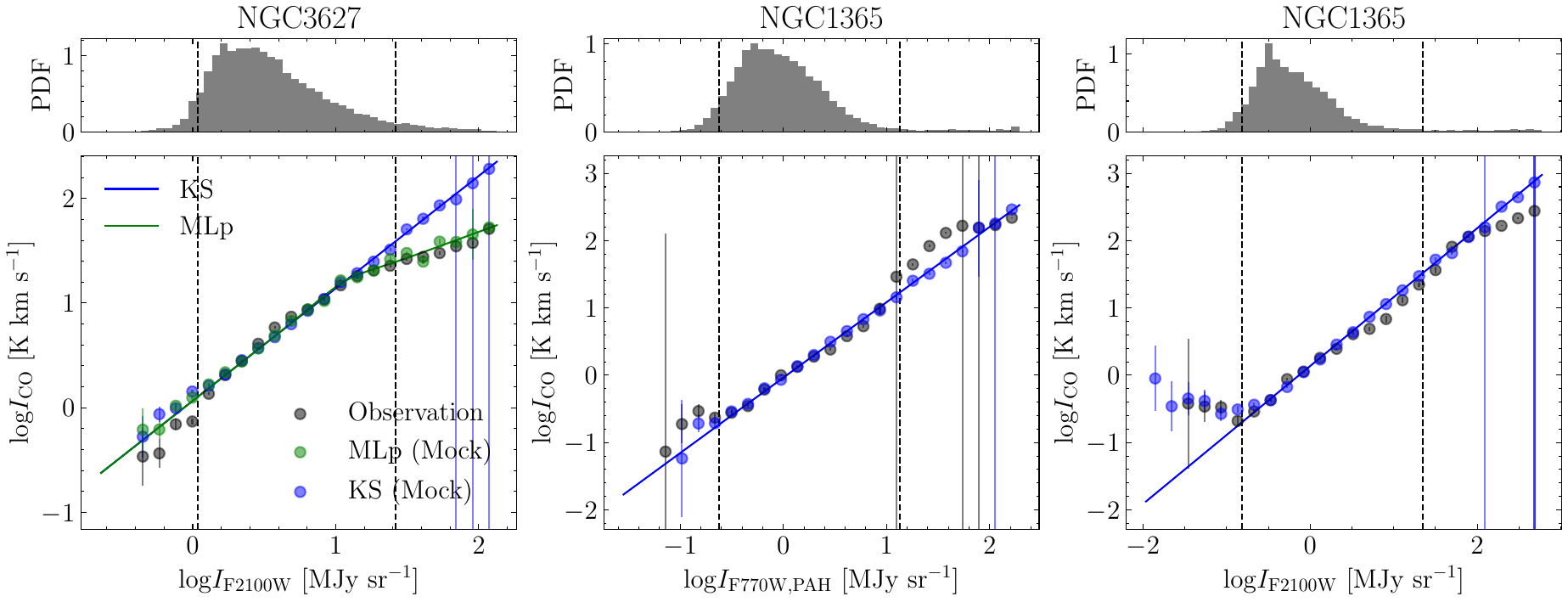}
    \caption{Examples of the three different cases of deviations from the log–log linear relation (case (a), (b), and (c), from left to right). In each panel, the lower sub-panel displays the logarithm of the median observed CO(2-1) flux in each observed $\redband$ or $\blueband$ flux bin as black circles (real data), blue circles (mock data generated by single log-log linear fitting), and green circles (mock data generated by piecewise log-log linear fitting, available only in the left panel). \edit{In the legend, ``KS'' denotes the results of single log-log linear fitting based on \ksmethod{}, while ``MLp'' denotes the results of piecewise log-log linear fitting based on \mlmethod{}.} Error bars represent the 1$\sigma$ uncertainty of the median. The best-fit relations for the single and piecewise log-log linear fits are shown as blue and green lines, respectively. The two vertical dotted lines mark the 5\% and 95\% percentiles of the observed $\redband$ or $\blueband$ flux distribution. These lines are also shown in the top sub-panel, where the histogram illustrates the marginalized distribution of the observed MIR bands. Note that data points with negative MIR bands measurements are excluded when plotting the marginalized distribution in logarithmic space but are included when calculating the 5\% and 95\% percentiles and applying regression analysis.
    \label{fig:three_cases}}
\end{figure*}

\subsection{Application of \raddest\ to PHANGS galaxies}
\label{sec:dataselection}

As previously noted, we employ the \ksmethod\ within \raddest\ due to its strong performance with large sample sizes. To ensure statistical robustness, we exclude datasets containing fewer than approximately 100 data points. Such low-data regimes typically occur in regions with AGN-like ionization conditions or in resolution-downgraded samples (see \autoref{sec:diff_spatial_scale} for details). We confirm that applying a more stringent sample size threshold (e.g., $\ge 1000$ data points) does not alter the key results of this study. The parameters $k$, $b$, and $\sigma$ derived from the \ksmethod\ are hereafter denoted as $\kKS$, $\bKS$, and $\sigKS$, respectively. 

\edit{We note that the parameter uncertainties reported by \raddest\ should be interpreted with caution. As demonstrated in \citetalias{Jing2024}, reliable uncertainty quantification remains a shared challenge among all widely used regression methods when applied to realistic log-log linear regression problems involving intrinsic scatter, measurement uncertainties in both variables, and complex data distributions. The \ksmethod\ tends to yield conservative uncertainty estimates, with confidence intervals that are generally broader than the true statistical uncertainties in validation tests, whereas the \mlmethod\ tends to produce overconfident uncertainty estimates that are systematically too small. Therefore, although we report the formal uncertainties returned by the regression procedure for completeness, our scientific interpretation primarily focuses on the best-fit parameter values and systematic trends rather than on the exact significance levels inferred from these uncertainties.}

In the \raddest\ framework, normalizing flows (NFs) are used to model the distributions $P(\bxerri|\bx)$, $P(\bx)$, and $P(\yerri|y)$ required for regression analysis. While the fitted NFs perform well for most of our dataset, they fail to generate mock data that pass the \edit{\ctest{}} in approximately 10\% of cases, even after extensive hyperparameter tuning. This failure primarily stems from overestimated uncertainties in the dataset. Specifically, we observe that within given uncertainty bins ($\err{\bx}$ or $\err{y}$, where $\bx$ corresponds to $\Iblue$, $\Igreen$, or $\Ired$, and $y$ corresponds to $\ICO$), the standard deviation ($\sigma_{\rm x, obs}$ or $\sigma_{\rm y, obs}$) of the observed values ($\obs{\bx}$ or $\obs{y}$) is systematically smaller than the reported uncertainty values ($\sigma_{\rm x, obs} < \err{\bx}$; $\sigma_{\rm y, obs} < \err{y}$). This discrepancy likely originates from the initial data reduction process or error propagation methodology. Since reprocessing the raw data or implementing more sophisticated error propagation falls outside the scope of this work, we adopt a heuristic approach for these problematic cases: we scale $\err{\bx}$  or $\err{y}$ to satisfy $\sigma_{\rm x, obs} = \err{\bx}$ or $\sigma_{\rm y, obs} = \err{y}$. This adjustment enables the NFs to pass the \edit{\ctest{}} in most instances. The few cases that still fail after scaling are excluded from subsequent analysis. We confirm that all key results presented in this study remain consistent regardless of whether we restrict analysis to non-problematic data or include all data without implementing this correction.

\begin{figure*}[ht!]
    \centering
    \includegraphics[width=0.95\textwidth]{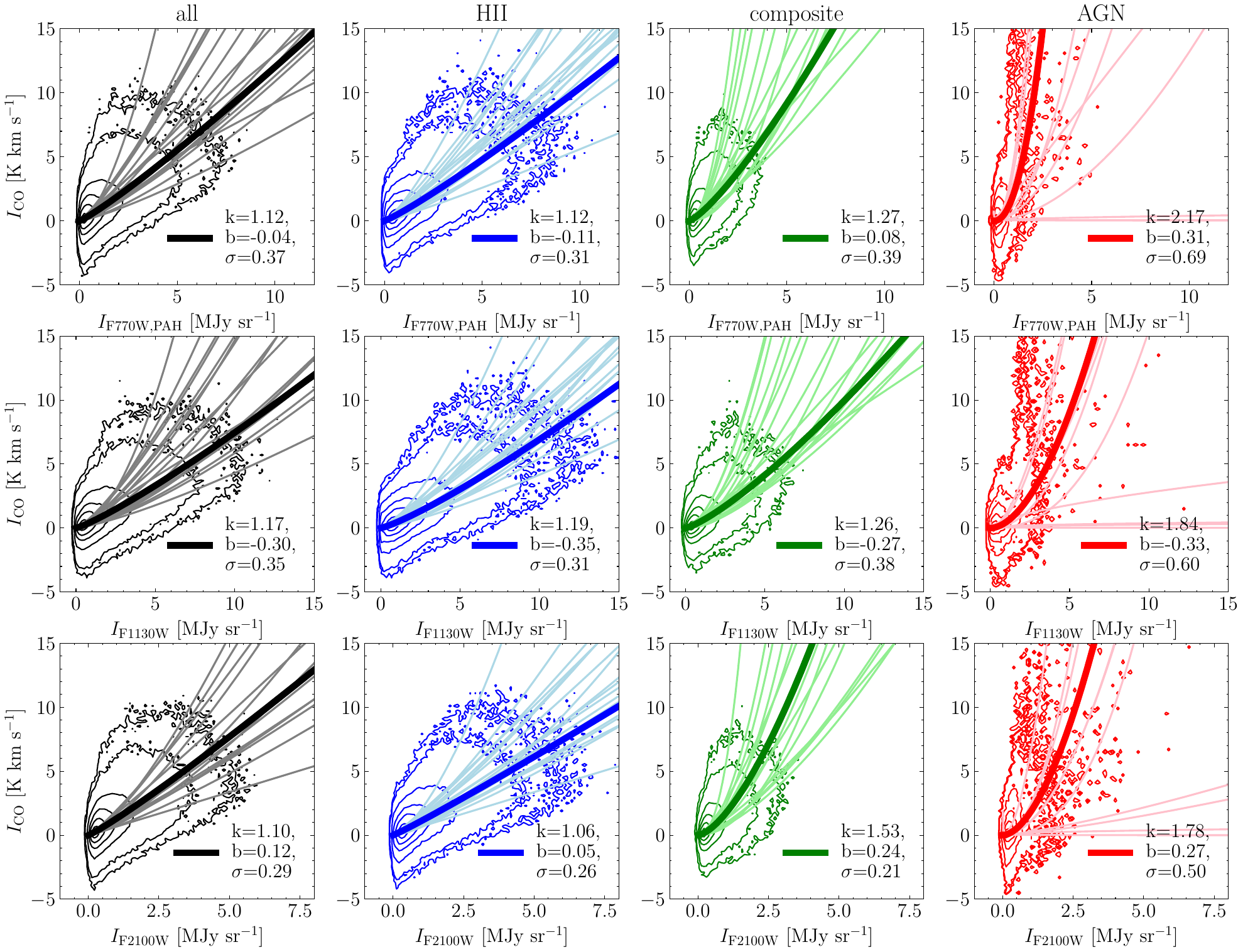}
    \caption{Scaling relation between $\ICO$ and $\Iblue$, $\Igreen$, and $\Ired$ (top to bottom) in different ionization conditions (all regions, \HII-like regions, composite-like regions, and AGN-like regions from left to right). In each panel, the contour showcase the distribution of spaxels from all galaxies, and the bold line is corresponding best-fit result. The thin lines are best-fit results for each galaxy.
    \label{fig:all_vs_each}}
\end{figure*}

The application of \raddest\ for regression analysis confirms that the log-log linear relation in~\autoref{eq:loglinear} provides a good description for the majority of data points ($\ge 80\%$ in most cases) across all galaxies in our sample and under all ionization conditions. Nevertheless, deviations from this relationship are detectable in most cases. We categorize these deviations visually into three types: 
\begin{enumerate}[label=(\alph*)]
  \item A deviation in the brightest regions that is well-described by a second log-log linear relation with a different slope. The ``no-deviation'' scenario is a special case of this, where the slope for the brightest regions is identical to that of the main trend.
  \item A deviation in the brightest regions that cannot be adequately fit by another log-log linear relation.
  \item A complex deviation that is more intricate than cases (a) and (b) and cannot be captured by simple functional forms. 
\end{enumerate}
Examples of these three cases are presented in~\autoref{fig:three_cases}. We note that, in addition to the three cases of deviation, the faint end usually presents deviations to varying degrees as well. This is due to the limited S/N in the faintest regions: the binning of data points is driven mainly by noise rather than by their true values, so the correlations are smeared out. This behavior is well reproduced by our generative model when uncertainties in both the x- and y-axes are taken into account. For instance, as shown in \autoref{fig:three_cases}, the median of mock data, represented as blue circles, aligns closely with that of real data, depicted as gray circles, in faint end bins.

In cases (a) and (b), a turning point $x_0$, defined as the MIR bands value above which the deviation occurs, can be identified. For case (a), this turning point is determined using a piecewise log-log linear formula:  
\begin{equation} \label{eq:p_loglinear}
    \begin{split}
        \log \ICO = & 
        \begin{cases}
        k_0 \log I_X + b_0 + \epsilon, & \text{if } I_X \leq x_0 \\
        k_1 \log I_X + b_1 + \epsilon, & \text{if } I_X > x_0
        \end{cases} \\
        & \epsilon \sim \mathrm{N}(0, \sigma^2)
    \end{split}
\end{equation}
where $k_0$, $b_0$, and $\sigma$ are fixed as $\kKS$, $\bKS$, and $\sigKS$, respectively. To fit $k_1$, $b_1$, and $x_0$, we apply the \mlmethod\ in \raddest, as this method demonstrates performance comparable to \ksmethod\ while being significantly more sensitive to minority data points that follow a distinct distribution compared to the majority. For case (b), the turning point is identified visually.

The strength of the deviation in cases (a) and (b) is quantified by calculating the difference of the median observed $\ICO$ between the real data points and the mock data points above the turning point, $x_0$. The mock data points are generated using the best-fit single log-log linear relationship. This difference is denoted as $\Delta_{\rm KS} = {\rm MEDIAN}(\obs{y}|\obs{x} > x_0) - {\rm MEDIAN}(\obs[,\rm gen]{y}|\obs[,\rm gen]{x} > x_0)$. For case (a), relative difference of the slope above and below the turning point ($R_{\Delta k} = (k_1 - \kKS)/\kKS$) is additionally available. 

\begin{figure*}[ht!]
    \centering
    \includegraphics[width=0.95\textwidth]{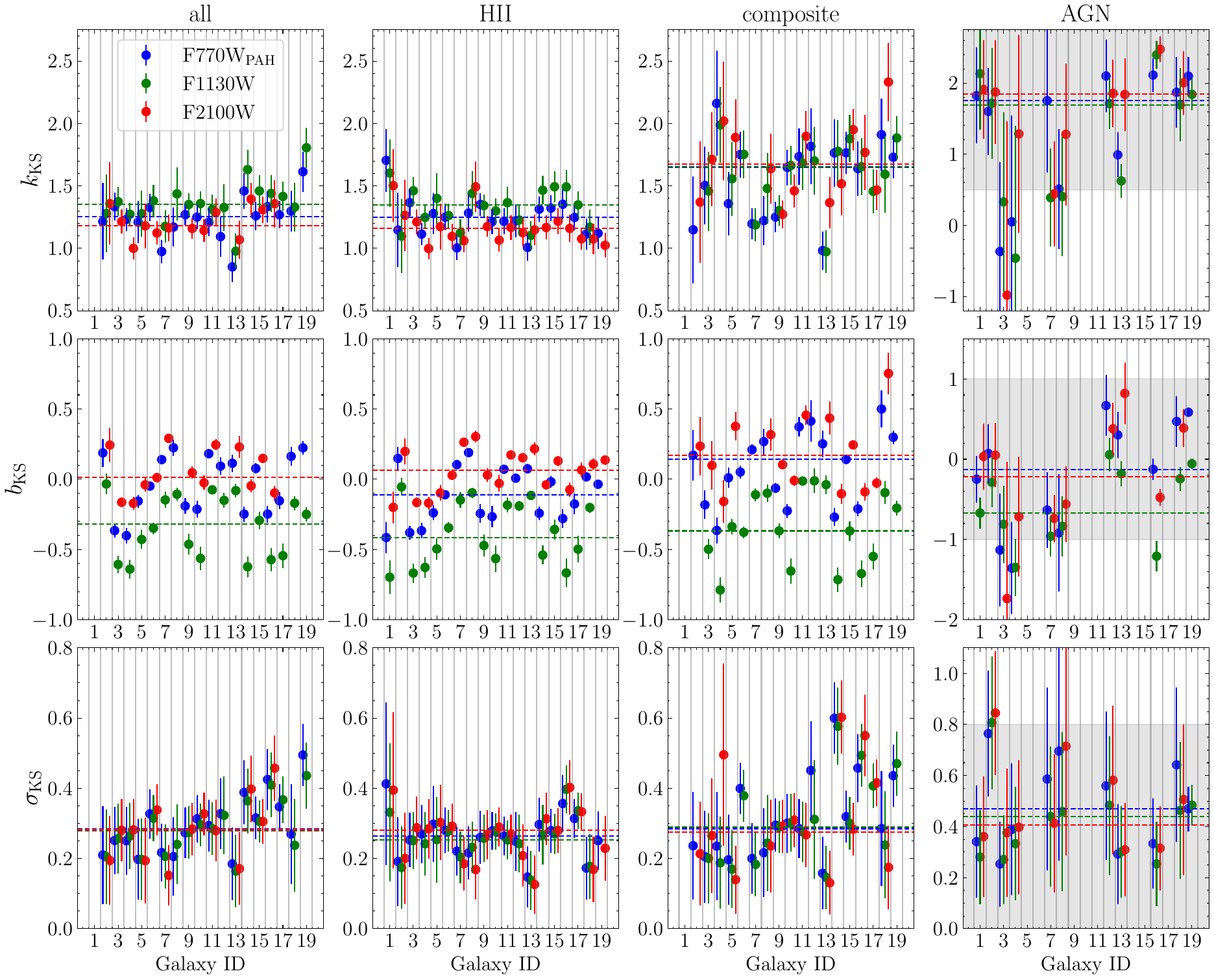}
    \caption{Best-fit slope $\kKS$, intercept $\bKS$, and intrinsic scatter $\sigKS$ based on \ksmethod\ for each galaxy, of different MIR bands (blue for $\Iblue$, green for $\Igreen$, and red for $\Ired$), and under different ionization conditions (all regions, \HII-like regions, composite-like regions, and AGN-like regions from left to right). In each panel, the galaxies are sorted on the x-axis by total stellar mass (and also by galaxy ID, as shown in \autoref{tab:galaxies}). The horizontal dashed lines with different colors show the median value of the best-fit parameters across different MIR bands. Error bars represent 1$\sigma$ uncertainties. In each row, panels in the first three columns share the same y-axis range, which is indicated as a shaded region in the last column panel.  
    \label{fig:kbs_scatter}}
\end{figure*}

\section{Result} \label{sec:result}

We first analyze the single log-log linear fits characterized by $\kKS$, $\bKS$, and $\sigKS$, examining their dependence on ionization condition in~\autoref{sec:ion_cond} and their variation from galaxy to galaxy in~\autoref{sec:res_each_galaxy}. We then examine the deviation from the log-log linear relation in~\autoref{sec:non_log_linear}. Finally, in~\autoref{sec:diff_spatial_scale}, we investigate the spatial scale dependence of the log-log linear relationship.

\begin{figure*}[ht!]
    \centering
    \includegraphics[width=0.95\textwidth]{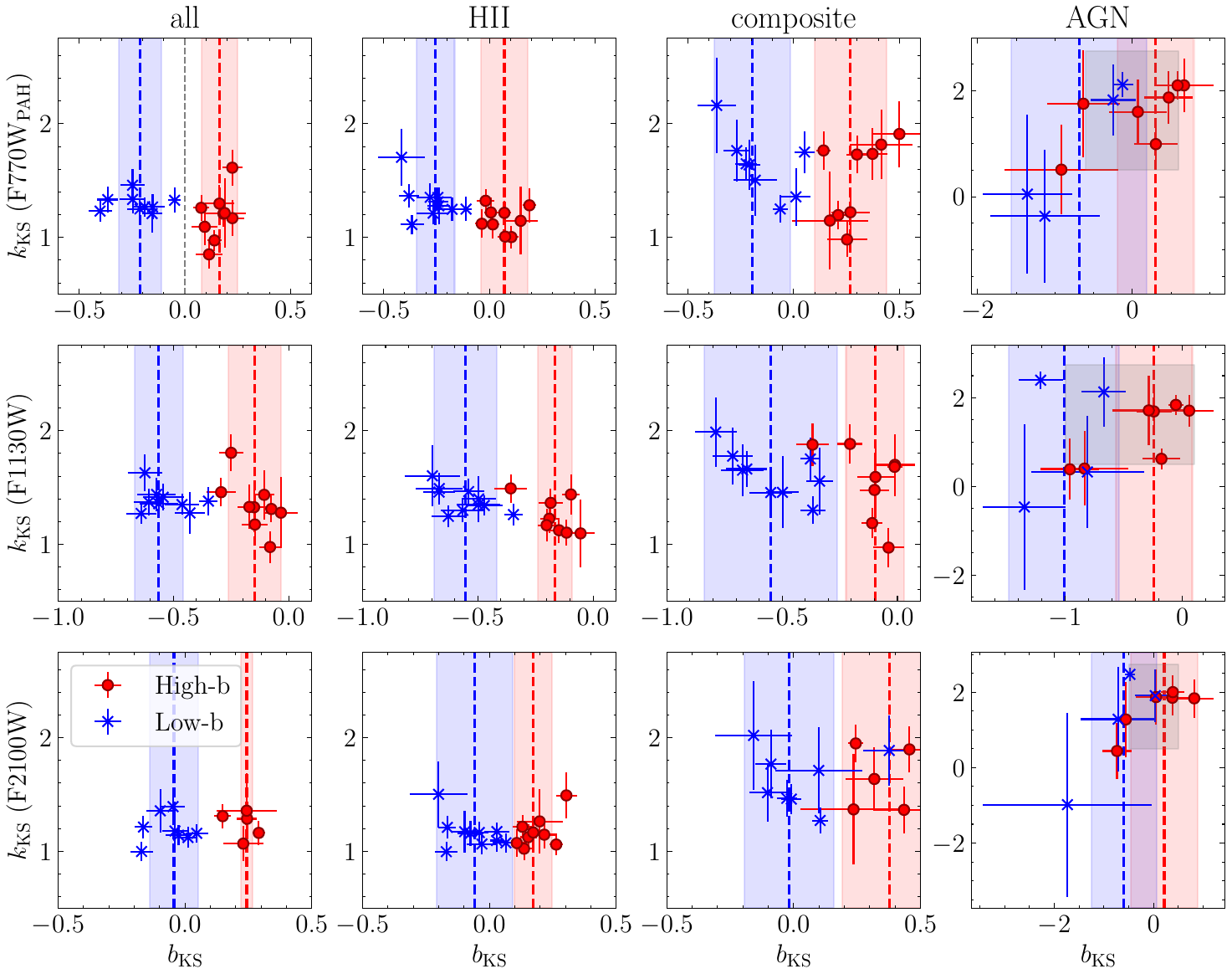}
    \caption{The correlation between the slope $\kKS$ and the intercept $\bKS$ is shown for each MIR band (top to bottom) and each ionization condition (left to right). The vertical dashed line in the top-left panel indicates $\bKS = 0$, the criterion used to classify galaxies into \highb\ and \lowb\ subgroups. The \highb\ galaxies are represented by red dots, while the \lowb\ galaxies are shown as blue crosses. In each row, the first three panels share identical x-axis and y-axis ranges; these ranges are highlighted as gray shaded regions in the last panel of each row. \edit{For each subgroup, the colored dashed lines denote the median $\bKS$, and the corresponding shaded regions indicate the median $\pm$ half-FWHM range estimated from the distribution of $\bKS$.}
    \label{fig:k_vs_b}}
\end{figure*}

\subsection{Dependence on Ionization Condition} \label{sec:ion_cond}

We present the correlations of $\ICO$ with $\Iblue$, $\Igreen$, and $\Ired$ obtained with all spaxels from all the 19 galaxies as black contours in the first column of \autoref{fig:all_vs_each}. The best-fit log-log linear scaling relations are shown as black thick lines, and the corresponding model parameters ($\kKS$, $\bKS$, $\sigKS$) are indicated in each panel. \edit{The corresponding best-fit parameters are presented in \autoref{app:res}.} Although we use a log-log linear formula, we display the correlations in linear space to avoid the clipping of data points with negative values caused by noise. Such clipping could create a misleading impression. The fitting techniques we employ properly handle these negative values in the log-log linear regression, thus ensuring unbiased estimations. As shown in these three panels, $\ICO$ remains correlated with $\Iblue$, $\Igreen$, and $\Ired$ on $\sim$ 100 pc scales. For the majority of data points, the correlations are well described by a single log-log linear relation. 
%This implies that the decoupling of CO from PAH and dust occurs at scales smaller than $\sim$100 pc.

In the second to fourth columns of \autoref{fig:all_vs_each}, we show the correlations and corresponding best-fit log-log linear relationships for spaxels classified as \HII-like, composite-like, and AGN-like regions. The correlations of $\ICO$ with $\Iblue$, $\Igreen$, and $\Ired$ differ significantly across different ionization conditions. \HII-like regions exhibit relatively flatter slopes $\kKS$, and resemble the correlations obtained when all ionization conditions are combined. This is consistent with the fact that all galaxies in our sample are main-sequence star-forming galaxies, where young stellar populations dominate the ionization. In composite-like regions, both $\ICO$ and the intensities of MIR bands show relatively narrower dynamical ranges, and the slopes $\kKS$ are steeper. In AGN-like regions, the dynamical range of $\ICO$ is broader, while that of the MIR bands is narrower, resulting in the steepest slopes $\kKS$ among the three ionization conditions. This may be related to the destruction of PAH and dust in AGN-like regions (see discussion in \autoref{sec:discuss_ion_cond}).

The scaling relations under different ionization conditions also exhibit varying $\bKS$ values. Overall, \HII-like regions have smaller $\bKS$ than composite-like and AGN-like regions. The comparison between composite-like and AGN-like regions depends on the MIR band: AGN-like regions have larger $\bKS$ than composite-like regions for $\blueband$, while the values are comparable for $\greenband$ and $\redband$.
% This implies that in harsh radiation filed dominated by AGN, the $\blueband$ carrier undergoes significantly stronger destruction. 

Regarding the intrinsic scatter $\sigKS$, we find the highest values in AGN-like regions. In the two PAH bands ($\blueband$ and $\greenband$), composite-like regions exhibit larger intrinsic scatter $\sigKS$ than \HII-like regions. In the dust band ($\redband$), however, \HII-like regions show larger intrinsic scatter $\sigKS$ than composite-like regions.

All the results reported above are based on the scaling relations of spaxels from all 19 galaxies. We will revisit the ionization dependence after analyzing galaxy-to-galaxy variations in the following subsection (\autoref{sec:ion_cond_in_gal_gal_var}).

\subsection{Galaxy-to-Galaxy Variation} \label{sec:res_each_galaxy}

The best-fit relations of individual galaxies are plotted as thin gray lines in \autoref{fig:all_vs_each}. Even at a fixed ionization condition, galaxy-to-galaxy variations persist, suggesting that this variation is driven by mechanisms beyond differences in ionization conditions. To provide more details on galaxy-to-galaxy variations, \autoref{fig:kbs_scatter} displays the best-fit $\kKS$, $\bKS$, and $\sigKS$ (from top to bottom) for different MIR bands (blue for $\Iblue$, green for $\Igreen$, and red for $\Ired$) across regions with different ionization conditions (from left to right) for each galaxy. \edit{The corresponding best-fit parameters are listed in \autoref{app:res}.} The galaxies are sorted by their total stellar mass, from low to high, as indicated by their IDs on the x-axis. As can be seen from the figure, while $\kKS$ (first row) and $\sigKS$ (third row) vary across galaxies, their deviations from the median values are within the 1–2$\sigma$ range. In contrast, $\bKS$ exhibits deviations exceeding 3$\sigma$, suggesting that galaxy-to-galaxy variation is primarily driven by variations in $\bKS$. 

\begin{figure}[t!]
    \centering
    \includegraphics[width=0.48\textwidth]{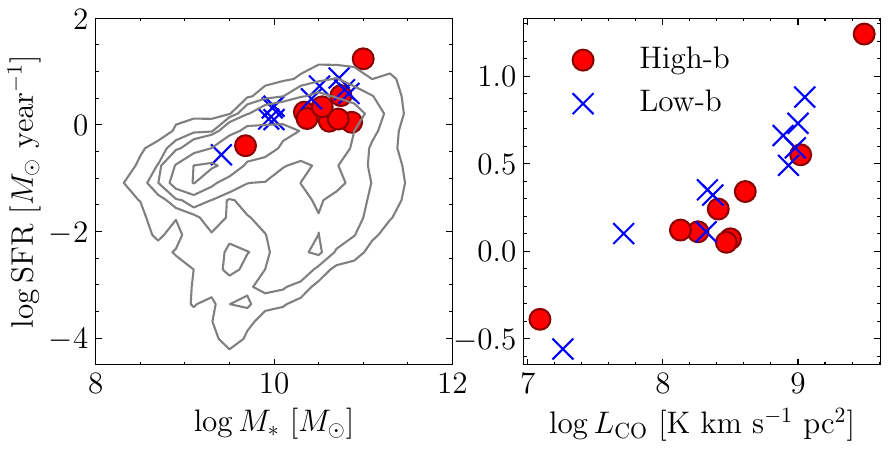}
    \caption{\textit{Left}: Distribution of low-$b$ (blue crosses) and high-$b$ (red circles) galaxies on $\log \SFR$ vs. $\log \Mstar$ diagram. The gray contour represents the distribution of a volume-limited sample of nearby galaxies, constructed from the MaNGA \citep{MaNGA_Bundy2015, MaNGA_Blanton2017, MaNGA_Wake2017} sample with galaxy weight corrections. \textit{Right}: Same as the left panel, but the x-axis is $\log \LCO$. 
    \label{fig:b_bimodality}}
\end{figure}

This is more clearly illustrated in \autoref{fig:k_vs_b}, where we show the $\kKS$ vs. $\bKS$ diagram for each ionization condition (columns) and each MIR bands (rows). Compared to their estimation uncertainties, $\bKS$ has a larger dynamical range than $\kKS$. The most striking feature of $\bKS$ is its bimodality, with two distinct subgroups of galaxies exhibiting higher and lower $\bKS$ values. Accordingly, we classify galaxies into \highb\ and \lowb\ subgroups based on the $\bKS$ value of the $\ICO$–$\Iblue$ scaling relation for spaxels under all ionization conditions (first panel of \autoref{fig:k_vs_b}). Galaxies with $\bKS > 0$ are categorized as \highb, while those with $\bKS < 0$ are categorized as \lowb{}\footnote{NGC5068 does not have successful fits under all ionization conditions because the NF fitting fails to pass the \edit{\ctest{}}. Therefore, it is classified as a \lowb{} galaxy based on the $\bKS$ value derived under \HII-like ionization conditions.}. These subgroups are marked in red (\highb) and blue (\lowb) in all panels of \autoref{fig:k_vs_b}. This classification clearly shows that the \highb\ (\lowb) galaxies consistently exhibit overall larger (smaller) $\bKS$ values across all MIR bands and ionization conditions, while their $\kKS$ values remain similar. 

\edit{To assess the significance of the bimodality, we calculate the median and FWHM of the $\bKS$ distributions for \highb{} and \lowb{} galaxies in each MIR band and ionization condition. We adopt non-overlapping FWHM ranges of the two subgroups as a criterion for clear separation of the bimodal components. According to this criterion, a clear bimodality is present in the all-spaxel sample and in \HII{}-like regions for all MIR bands, except for $\redband{}$ in \HII{}-like regions, where the FWHM ranges of the two subgroups show a slight overlap. In composite-like regions, the two subgroups remain distinguishable, but their separation is weaker than in the all-spaxel and \HII{}-like samples. In AGN-like regions, no clear bimodality is evident. }
% Overall, although \highb{} (\lowb{}) galaxies consistently exhibit larger (smaller) $\bKS$ values across all MIR bands and ionization conditions, the bimodality is most prominent in \HII{}-like regions, less pronounced in composite-like regions, and largely absent in AGN-like regions. Among the MIR bands, the separation is stronger in $\blueband{}$ and $\greenband{}$ than in $\redband{}$.

\subsubsection{Dependence on Ionization Condition} \label{sec:ion_cond_in_gal_gal_var}

Keeping the galaxy-to-galaxy variation in mind, we revisit the ionization condition dependence of the scaling relations. \edit{As can be seen in \autoref{fig:kbs_scatter}}, the parameter $\kKS$ still follows the same ascending order among \HII-like, composite-like, and AGN-like regions when comparing their median values, even after accounting for galaxy-to-galaxy variation. This is consistent with the conclusion that the galaxy-to-galaxy variation is not primarily driven by differences in $\kKS$. For $\bKS$, while differences are observed across different ionization conditions, these differences are significantly smaller than the galaxy-to-galaxy variation in $\bKS$. The median value of $\sigKS$ is comparable between \HII-like and composite-like regions, whereas AGN-like regions exhibit a consistently larger $\sigKS$. This suggests that the difference in $\sigKS$ between \HII-like and composite-like ionization conditions, as derived from spaxels across all 19 galaxies, is primarily due to the varying levels of galaxy-to-galaxy variation under different ionization conditions. However, AGN-like regions retain a larger $\sigKS$, even after removing the contribution from galaxy-to-galaxy variation.

\subsubsection{Comparison across Different MIR Bands}

It is also valuable to compare the scaling relations across different MIR bands in the context of galaxy-to-galaxy variation. \edit{As shown in \autoref{fig:kbs_scatter}, the median values of $\kKS$ within \HII-like regions exhibit a slight difference across MIR bands, increasing in the order of $\redband$, $\blueband$, and $\greenband$.} This order is consistently observed in individual galaxies. However, the magnitude of this difference is small compared to the galaxy-to-galaxy variation within each MIR band. In composite-like and AGN-like regions, the slope variation across different MIR bands is even less significant. Clear differences in $\bKS$ are observed across the three MIR bands in \HII-like regions, reflecting distinct overall $\ICO/\Iblue$, $\ICO/\Igreen$, and $\ICO/\Ired$ ratios. In composite-like and AGN-like regions, the $\bKS$ values for $\blueband$ and $\redband$ are more similar, while $\greenband$ shows a smaller $\bKS$. For $\sigKS$, there are no significant differences across the MIR bands within any given ionization condition. This suggests that the coupling between CO, PAH, and dust remains similarly tight at $\sim$100 pc scales (see \autoref{sec:discuss_diff_xbands} for detailed discussions).

\edit{An interesting feature is that, among the three MIR bands, $\blueband{}$ exhibits a behavior more similar to $\redband{}$, while $\greenband{}$ shows a more distinct trend. One possible explanation is the contribution of hot dust continuum emission to $\blueband{}$. However, spectral decomposition and modelling studies in the literature \citep{Whitcomb2023,Sutter2024_PAHinDust,Chown2025_PAHinMIR, Donnelly2025_PAHinMIR} suggest that the dust continuum contribution in $\greenband{}$ is comparable to, or even larger than, that in $\blueband{}$. Therefore, dust continuum contamination alone is unlikely to be the primary driver of this similarity. A definitive explanation requires detailed physical modelling; nevertheless, one plausible scenario is that smaller PAHs, which contribute more strongly to the 7.7 $\mathrm{\mu}$m feature than to the 11.3 $\mathrm{\mu}$m feature, preferentially exist in warmer environments where stronger hot dust emission traced by $\redband{}$ is also present.}

% Nevertheless, the significant bimodality can only be detected for $\bKS$ of $\ICO$-$\Iblue$ and $\ICO$-$\Igreen$ correlations on spaxels with all, \HII-like, and composite-like ionization conditions. 

\subsubsection{Dependence on Galaxy Properties}

\begin{figure*}[ht!]
    \centering
    \includegraphics[width=0.9\textwidth]{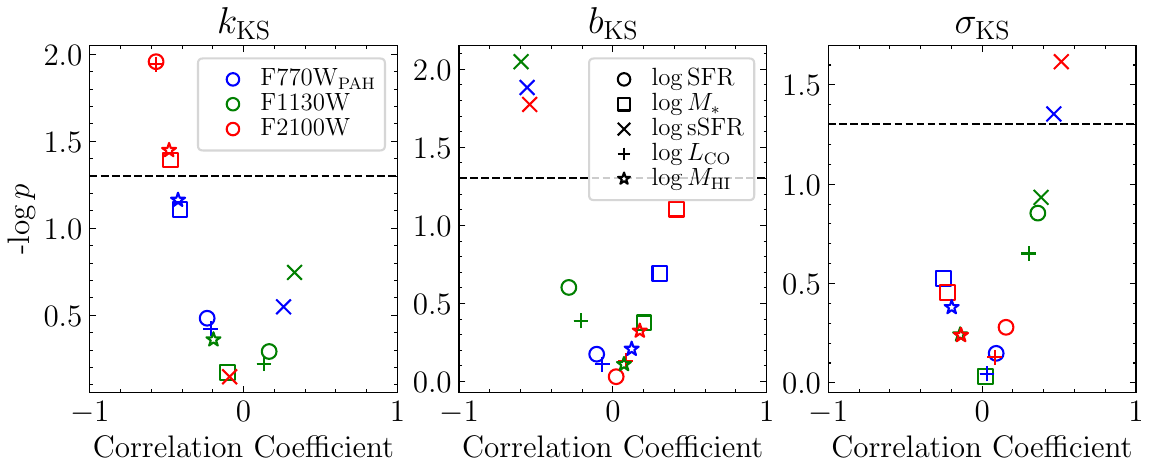}
    \caption{Correlation coefficient and corresponding $-\log p$ between the best-fit parameters ($\kKS$, $\bKS$, and $\sigKS$, from left to right) and various global properties ($\log \SFR$, $\log \Mstar$, $\log \sSFR$, $\log \LCO$, and $\log \MHI$, represented by different symbols) for different bands ($\blueband$, $\greenband$, and $\redband$, shown in blue, green, and red, respectively). Horizontal dashed line in each panel represents $-\log p \approx 1.3$ ($2\sigma$). 
    \label{fig:vs_property_HII}}
\end{figure*}

\begin{figure*}[ht!]
    \centering
    \includegraphics[width=0.9\textwidth]{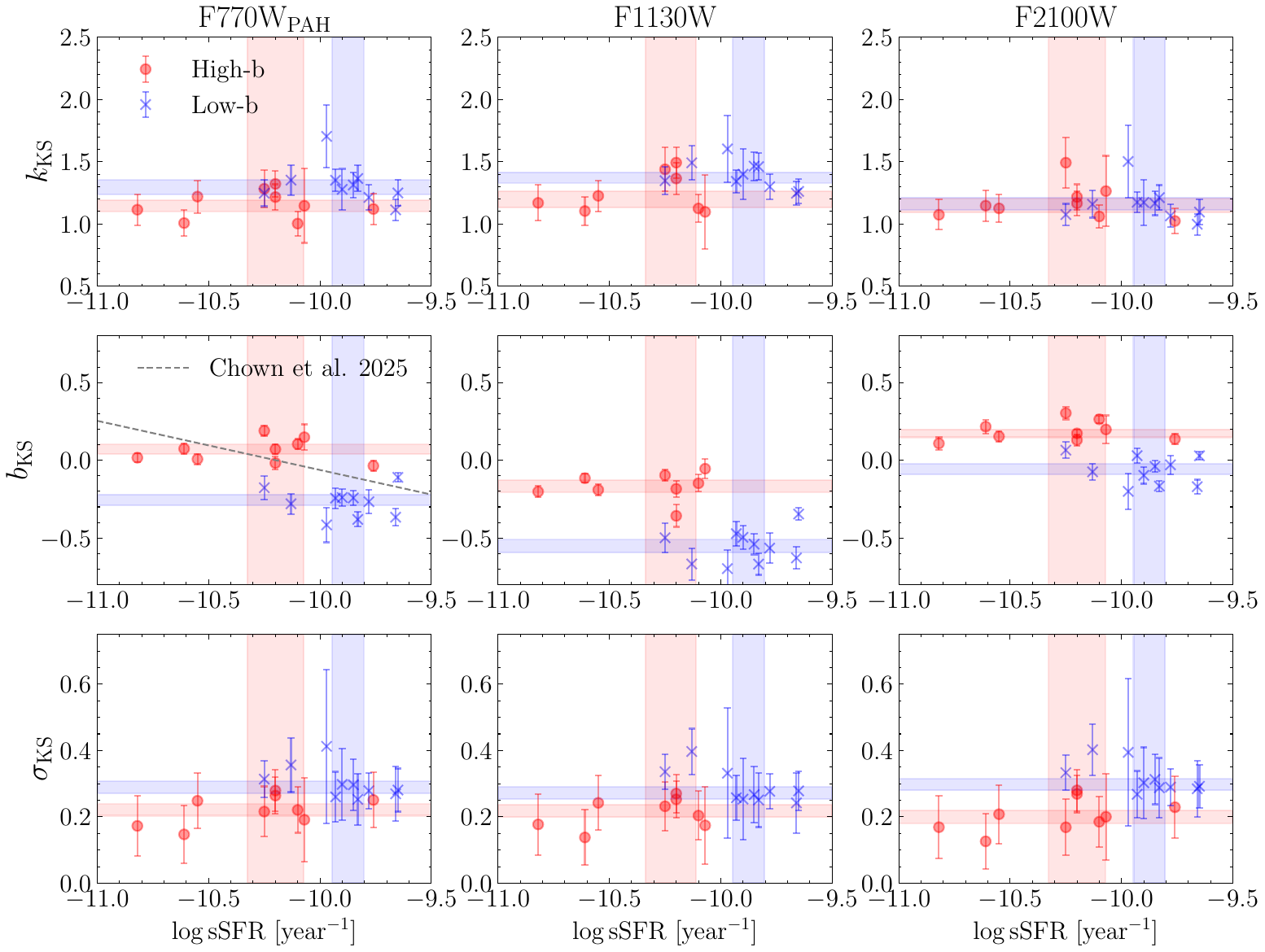}
    \caption{Correlation between the best-fit parameters and galactic $\log \sSFR$ for different MIR bands ($\blueband$, $\greenband$, and $\redband$ from left to right). High-$b$ (low-$b$) galaxies are shown as red circles (blue crosses), with error bars representing 1$\sigma$ uncertainties. The correlation between $b$ and $\log \sSFR$ reported by \cite{Chown2024} is displayed as a gray dashed line in the corresponding panel (left panel in the second row). \edit{The red (blue) horizontal and vertical shaded regions indicate the 1$\sigma$ confidence intervals of the median x- and y-axis values for \highb{} (\lowb{}) galaxies.}
    \label{fig:vs_sSFR_HII}}
\end{figure*}

\edit{As mentioned earlier, galaxies with higher (or lower) $\bKS$ values in one ionization condition and MIR band also tend to exhibit higher (or lower) $\bKS$ values in other ionization conditions and MIR bands. It implies that the physics driving the $\bKS$ bimodality regulates the correlation between CO, PAH, and dust in entire galaxies. To explore the origin of this bimodality, we examine the global properties of \lowb\ and \highb\ galaxies, including star formation rate ($\SFR$), stellar mass ($\Mstar$), and CO luminosity ($\LCO$). As shown in \autoref{fig:b_bimodality}, although all galaxies lie on the star-forming main sequence, \lowb\ galaxies tend to have higher $\SFR$ at a given $\Mstar$ or $\LCO$. This may be related to enhanced PAH and dust emission caused by stronger overall UV background in galaxies with stronger star formation activity (see \autoref{sec:discuss_gal_gal_var} for discussion).}

\edit{Next, we systematically investigate whether the galaxy-to-galaxy variation of the best-fit $\kKS$, $\bKS$, and $\sigKS$ can be explained by differences in their global properties.} For this analysis we focus on \HII-like regions due to their well-defined physical origins and the availability of a sufficient number of data points. In contrast, composite-like regions exhibit complex physical origins, while AGN-like regions are limited by insufficient data, leading to noisier results. We calculate the correlation coefficients and corresponding $p$-values between the best-fit parameters ($\kKS$, $\bKS$, and $\sigKS$) and various global properties ($\log \SFR$, $\log \Mstar$, $\log \sSFR$, $\log \LCO$, and $\log \MHI$) for different bands ($\blueband$, $\greenband$, and $\redband$). The results are presented in \autoref{fig:vs_property_HII}. 

For $\kKS$, anti-correlations at the 2$\sigma$ level are observed in $\redband$ with $\log \SFR$, $\log \Mstar$, $\log \LCO$, and $\log \MHI$. This suggests a possible role for these global properties in modulating the correlation between $\ICO$ and $\Ired$. However, due to the low statistical significance, these correlations cannot be confirmed with the current dataset. Future analyses using larger samples are required to either confirm or refine these results.

\edit{For $\bKS$ and $\sigKS$, $2\sigma$ correlations with $\log \sSFR$ are evident across all bands, except for $\sigKS$ in $\greenband$. This is consistent with the relationship between $\log \sSFR$ and $\bKS$ in $\blueband$ reported by \cite{Leroy2023_1kpc} and \cite{Chown2024}.} As shown in \autoref{fig:vs_sSFR_HII}, however, these correlations primarily arise from the bimodality of intercept $\bKS$: \edit{\highb\ galaxies exhibit lower median $\log \sSFR$ and $\sigKS$ values than \lowb\ galaxies.} Within each subgroup (\highb\ or \lowb), these dependences weaken significantly. This suggests that the underlying physics that drive the intercept bimodality play a more fundamental role than apparent $\sSFR$.

\begin{figure*}[ht!]
    \centering
    \includegraphics[width=0.95\textwidth]{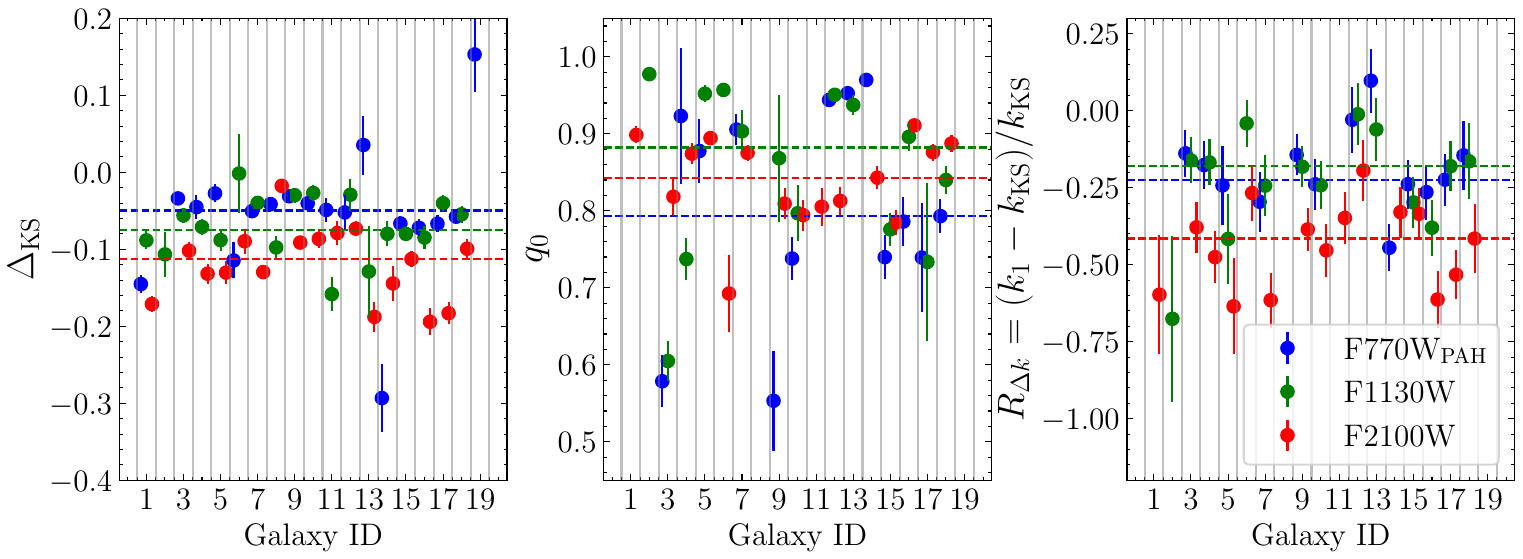}
    \caption{Same as \autoref{fig:kbs_scatter}, but showing the difference of the median observed $\ICO$ between the real data points and the mock data points above the turning point ($\Delta_{\rm KS}$), the quantile of the turning point ($q_0$), and the relative difference between the slope above and below the turning point ($(k_1 - \kKS)/\kKS$), in \HII-like regions of each galaxy. $\Delta_{\rm KS}$ and $q_0$ are available for both cases (a) and (b), while $(k_1 - \kKS)/\kKS$ is only available for case (a).
    \label{fig:non_loglinear_scatter}}
\end{figure*}

\subsection{Deviations from the Log-log Linear Relation} \label{sec:non_log_linear}

In this subsection, we analyze the deviations from the log–log linear relation observed in the brightest regions. Again, we focus on \HII-like regions due to their clear physical origins and the availability of a sufficient number of data points. For \HII-like regions, the fractions of cases (a), (b), and (c) are 75\%, 20\%, and 5\%, respectively (refer to \autoref{sec:regression} for classification and description of these cases). We concentrate on cases (a) and (b) since they are the dominant cases, and their deviations from the log–log linear relation can be characterized using well-defined statistical methods. As described in~\autoref{sec:dataselection}, we quantify the strength of the deviation by two parameters: $\Delta_{\rm KS}$ (the difference of the median observed $\ICO$ between the real data points and the mock data points above the turning point) and $R_{\Delta k}$ (the relative difference of the slope above and below the turning point). In addition, we consider a third parameter: the quantile of the turning point $q_0$, which is defined as the fraction of data points with the considered MIR intensity below the turning point. 

\autoref{fig:non_loglinear_scatter} displays (from left to right) $\Delta_{\rm KS}$, $q_0$ and $R_{\Delta k}$ as function of galaxy ID, for the three MIR bands as indicated. As shown, in all three bands, all galaxies exhibit negative $\Delta_{\rm KS}$ values or values that are statistically consistent with zero. This indicates that, where deviations exist, the observed values are systematically smaller than the single log-log linear predictions in the brightest regions. The only exception is the $\blueband$ of NGC1365, as shown in the middle panel of \autoref{fig:three_cases}. This discrepancy is likely related to the presence of bar-induced inflows, a central starburst, and a powerful AGN in that galaxy. However, the exact causes require further analysis. When comparing the three MIR bands, we see more negative values of $\Delta_{\rm KS}$ in the dust band ($\redband$) than in the PAH bands ($\blueband$ and $\greenband$), indicating that the strongest deviation occurs in $\redband$. 

The quantile of turning points, $q_0$, shows a wide distribution between $0.5$ and $1.0$ (center panel of \autoref{fig:non_loglinear_scatter}). This result shows that the majority of data points (at least $>50\%$, $>80\%$ in most case) can be well described by a single log-log linear relation in all galaxies. On the other hand, however, the wide range of $q_0$ indicates that although deviations consistently occur in brighter regions, there is no characteristic value of $q_0$. Similarly, an examination of the absolute turning point value, $x_0$, also reveals no characteristic value. These findings suggest that PAH or dust intensity is not the primary driver of the deviation from the log-log linear relation. It is noticeable that, the dust band exhibits a narrower distribution of $q_0$ than two PAH bands. 

As shown in the right panel of \autoref{fig:non_loglinear_scatter}, 
the relative difference in the slope, $R_{\Delta k}$, is smaller than zero in all bands, with the overall largest slope differences observed in the dust band. This result indicates that, when deviations occur, the slope above the turning point is consistently flatter. This result aligns with the negative values of $\Delta_{\rm KS}$ and suggests that the physical mechanism driving the deviation likely suppresses CO(2-1) emission and/or enhances PAH and dust emission.

\begin{figure}[t!]
    \centering
    \includegraphics[width=0.45\textwidth]{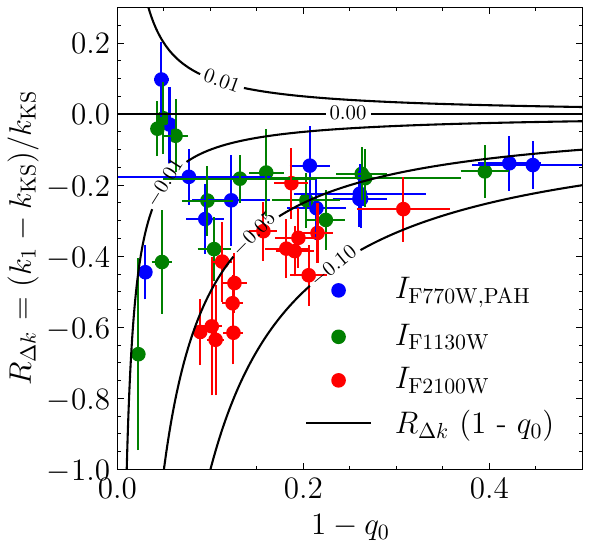}
    \caption{Diagram of $R_{\Delta k} = (k_1 - \kKS)/\kKS$ vs. $1 - q_0$. Results using $\Iblue$, $\Igreen$, and $\Ired$ as MIR bands are shown in blue, green, and red, respectively, with corresponding 1$\sigma$ uncertainties as error bars. The solid black lines represent different levels of $R_{\Delta k}(1 - q_0)$.
    \label{fig:dk_vs_q0}}
\end{figure}

The combination of $q_0$ and $R_{\Delta k}$, expressed as $R_{\Delta k}(1 - q_0)$, provides a more comprehensive description of the deviation. Specifically, $1 - q_0$ represents the fraction of data points above the turning point, i.e. those that cannot be described by the single log-log linear relation, while $R_{\Delta k}$ quantifies the degree of deviation for these data points. \autoref{fig:dk_vs_q0} displays our galaxies on the plane of $R_{\Delta k}$ versus $1-q_0$, with points of different colors for the three MIR bands. The black lines represent constant values of $R_{\Delta k}(1 - q_0)$ as indicated. As can be seen, the two PAH bands ($\blueband$ and $\greenband$) exhibit similar distributions on this diagram, with larger dynamical ranges in both $1 - q_0$ and $R_{\Delta k}$, along with lower absolute values of $R_{\Delta k}(1 - q_0)$, when compared to the dust band ($\redband$). In contrast, the dust band shows a more concentrated distribution of $1 - q_0$ and $R_{\Delta k}$, as well as higher absolute values of $R_{\Delta k}(1 - q_0)$, thus clearly separated from the PAH bands in this diagram. This result highlights the stronger deviation and flattening in the dust band than in the PAH bands as noticed from the previous figure. 
%This suggests that the deviation cannot be entirely attributed to the suppression of CO(2-1) emission in the brightest regions; otherwise, a similar deviation pattern would be observed across all three MIR bands analyzed. Furthermore, 
This result implies that the physical origin of the deviation likely works differently between the PAH and dust components  (see \autoref{sec:discuss_non_log_linear} for discussion on plausible mechanisms behind this result).

\begin{figure*}[ht!]
    \centering
    \includegraphics[width=0.95\textwidth]{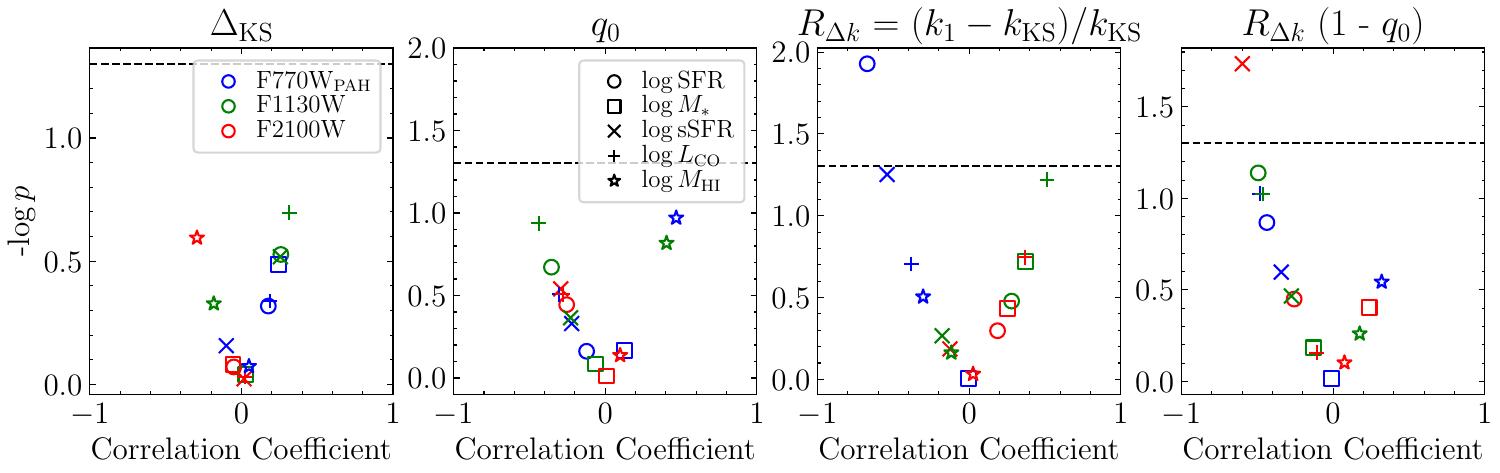}
    \caption{Same as \autoref{fig:vs_property_HII}, but for $\Delta_{\rm KS}$, $q_0$, $R_{\Delta k}$, and $R_{\Delta k}(1 - q_0)$.
    \label{fig:nonlinear_vs_property_HII}}
\end{figure*}

\begin{figure*}[th!]
    \centering
    \includegraphics[width=0.95\textwidth]{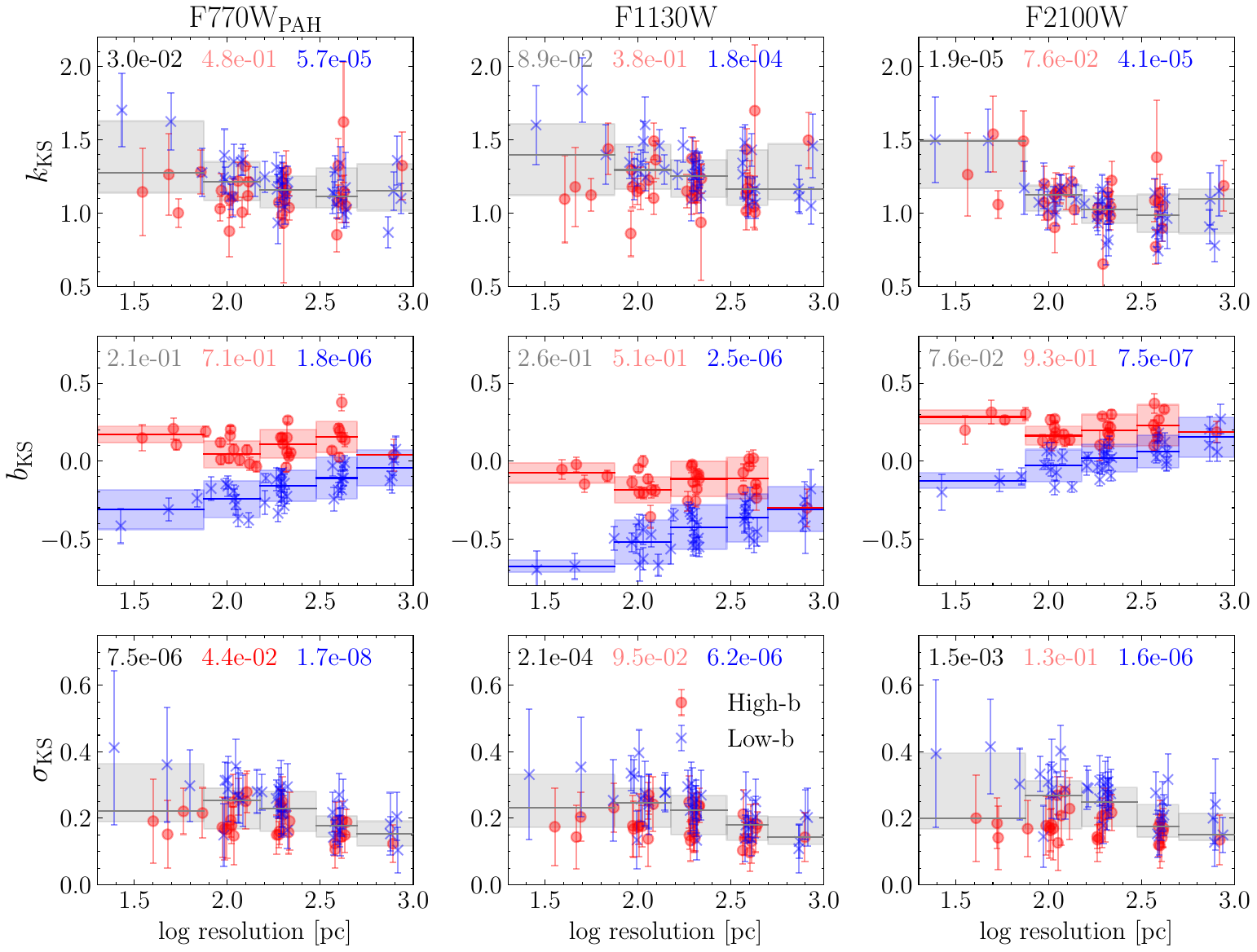}
    \caption{Best-fit slope $\kKS$, intercept $\bKS$, and intrinsic scatter $\sigKS$ in \HII-like regions at different spatial scales (resolutions) for different MIR bands ($\blueband$, $\greenband$, and $\redband$ from left to right). The low(high)-$b$ galaxies are represented as blue crosses(red circles), with error bars indicating 1$\sigma$ uncertainties. \edit{The gray horizontal lines and shaded regions in the first and third rows indicate the median values and 16\%--84\% intervals of the distributions in each resolution bin. In the second row, the red and blue horizontal lines denote the medians for \highb{} and \lowb{} galaxies in each resolution bin, respectively, and the shaded regions indicate the corresponding FWHM ranges.} \edit{The $p$-values from Pearson correlation tests for all galaxies, \highb\ galaxies, and \lowb\ galaxies are shown at the top of each panel in black, red, and blue, respectively. Text is displayed with reduced opacity when $p > 0.05$, indicating correlations that are not statistically significant at approximately the 2$\sigma$ level.}
    \label{fig:vs_resolution_HII}}
\end{figure*}

\begin{figure*}[ht!]
    \centering
    \includegraphics[width=0.95\textwidth]{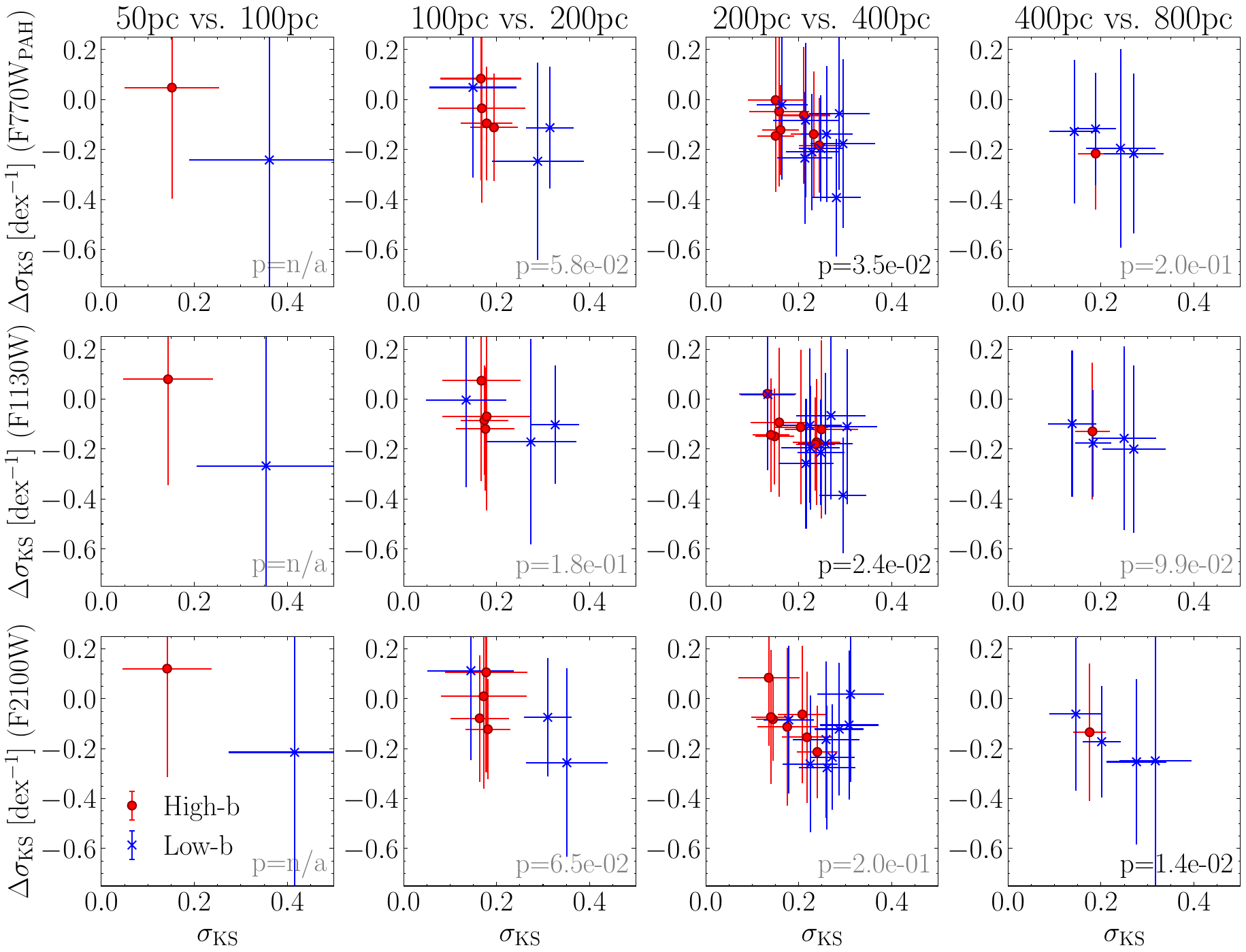}
    \caption{Difference in $\sigKS$ estimated in \HII-like regions between relatively larger spatial scale  (lower-resolution) and smaller spatial scale (higher-resolution) data (calculated as the former minus the latter), plotted as a function of $\sigKS$ of relatively smaller scales. The low(high)-$b$ galaxies are represented as blue crosses(red circles), with error bars indicating 1$\sigma$ uncertainties. \edit{The $p$-values from Pearson correlation tests are shown in the lower right corner of each panel. Text is displayed with reduced opacity when $p > 0.05$, indicating correlations that are not statistically significant at approximately the 2$\sigma$ level.}
    \label{fig:Dsigma_vs_sigma}}
\end{figure*}

In \autoref{fig:nonlinear_vs_property_HII}, we examine the correlation between global galaxy properties and the parameters that describe the degrees of deviation from the log–log linear relation, as quantified by both $p$-values and correlation coefficients. We identify only two 2$\sigma$ anti-correlations: one between $(k_1 - \kKS)/\kKS$ and $\log \SFR$, and another between $R_{\Delta k}(1 - q_0)$ and $\log \sSFR$. These findings suggest that the global properties explored here are unlikely to be the primary drivers of the deviation from the log-log linear scaling relation.

\subsection{Scaling Relations at Different Spatial Scale} \label{sec:diff_spatial_scale}
% The "resolution" are replaced by "Spatial Scale" to emphasize the analysis on lower resolution data shoule be not only considered as the mimic of observation effect but also the tracer of mechanism that regulate the scaling relation at different scale.

To investigating the variation of the scaling relation across different spatial scales, we reduce the spatial resolution of the data to 50, 100, 200, 400, and 800 pc through Gaussian convolution. Additionally, we resample the pixel size to half of the PSF size to minimize correlations between data points. For a given galaxy, the spatial resolution can only be downgraded to a value comparable or poorer than its original resolution after preprocessing. Consequently, not all of the listed resolutions are available for all galaxies. Furthermore, resolution downgrading and subsequent resampling reduce the number of data points. Samples with fewer than $\sim 100$ data points are excluded, as described in \autoref{sec:phangs}. Since our galaxies are dominated by \HII-like regions and these regions exhibit the highest completeness (i.e., they are less likely to be excluded due to limited sample sizes after resolution downgrading) across different resolutions, we focus on \HII-like regions in this analysis.

\autoref{fig:vs_resolution_HII} shows the best-fit model parameters ($\kKS$, $\bKS$, $\sigKS$) as function of physical scale, with red and blue symbols for \HII\ regions in \highb\ and \lowb\ galaxies. \edit{As can be seen, $\kKS$ shows little dependence on spatial scale for scales larger than $\sim100$ pc in all bands. However, $\kKS$ increases at the smallest spatial scale ($\le 100$ pc), reaching higher values than those measured at larger scales. The increase is most pronounced for $\Ired$. The higher values of $\kKS$ at small spatial scales suggest that the coupling between CO, PAH, and dust is primarily associated with their compact components rather than the diffuse ones.}

For $\bKS$, we observe different effects of spatial scale on \lowb\ and \highb\ galaxies. Specifically, $\bKS$ for \highb\ galaxies remains nearly constant across different spatial scales, whereas $\bKS$ for \lowb\ galaxies increases with increasing scale. This trend makes the bimodality unrecognizable in low-resolution data.

Overall, $\sigKS$ decreases with increasing spatial scale (decreasing resolution). This trend qualitatively aligns with the expectation that random scatter cancel out as the PSF covers larger regions. However, this random scatter cancellation is not the complete explanation. On the one hand, $\sigKS$ is defined in logarithmic space, while flux accumulation within the PSF occurs in linear space. On the other hand, neighboring regions are likely physically correlated. Therefore, the cancellation mechanism only applies to scatter caused by physical processes operating on scales smaller than the PSF of the higher resolution data. 
Notable differences in spatial scale dependence of $\sigKS$ are also apparent between \lowb\ and \highb\ galaxies. \edit{This distinction may be partly driven by the fact that \highb\ galaxies tend to have lower $\sigKS$, while the reduction in $\sigKS$ during resolution downgrading appears weaker when $\sigKS$ is already small. In this interpretation, the observed difference would not necessarily imply an intrinsic distinction between \lowb\ and \highb\ galaxies.} \edit{Evidence for this interpretation is provided in \autoref{fig:Dsigma_vs_sigma}}, where we plot the difference of $\sigKS$ between different spatial scales ($\Delta \sigKS$) as a function of $\sigKS$ at relatively small spatial scales for each galaxy. \edit{A tendency for $\Delta \sigKS$ to decrease with increasing $\sigKS$ is seen in most panels and across all bands, although the statistical significance of this trend is generally modest.} Moreover, at a given $\sigKS$, both \lowb\ (blue crosses) and \highb\ (red circles) galaxies exhibit similar $\Delta \sigKS$.

It is important to note that the results obtained at larger spatial scales (lower resolution) are not solely due to observational effects. They also provide insights into the mechanisms regulating scaling relations across different spatial scales, which are discussed in \autoref{sec:discuss_diff_scale}.

\section{Discussion} \label{sec:discussion}

\subsection{Comparison to Previous Work} \label{sec:compare_to_literatures}

\subsubsection{MIR fluxes from WISE}

The WISE W3 band, centered at $\sim 12\ \mu{\rm m}$, is broadly comparable to the JWST F1130W filter. The correlation between CO(1-0) emission and W3-band surface brightness was investigated in the EDGE-CALIFA sample by \cite{Chown2021}, who found a mean slope of $\sim$1 at kpc scales. This is slightly flatter than the mean slope we find for F1130W at 800 pc scale, $\langle\kKS\rangle\sim 1.2$ (top central panel of \autoref{fig:vs_resolution_HII}). The discrepancy likely stems from the different CO transitions used. Indeed, a slightly steeper correlation for CO(2-1) with W3 than for CO(1-0) with W3 has been reported in other studies \citep{Gao2019, Leroy2023_1kpc}, where CO(1-0) versus W3 relations yield slopes consistent with \cite{Chown2021} \citep{Gao2019, Gao2022, Leroy2023_1kpc}. 

Furthermore, \citet{Chown2021} investigated the spatial scale dependence of the CO–W3 relation by comparing contiguous regions of increasing size. They found that while the slopes were consistent from kpc scales up to entire galaxies, the scatter in the relation decreased as the region size approached that of a whole galaxy (their Figure 8). This led the authors to suggest that the global correlation observed at galaxy scales emerges from the local correlation operating at kpc scales. Comparing our results at the 800 pc scale with those reported for galactic scales in the literature confirms the consistency of slopes from kpc to galaxy-wide scales. However, by extending the analysis to sub-kpc resolution, we uncover a key new insight: significant variation in $\kKS$ that depends on local physical conditions (e.g., ionization condition dependences and the deviations from the log–log linear relation in the brightest regions). This indicates that the local correlation is intrinsically different from the global one, even though the latter is an aggregate of the former. In other words, the correlation is modulated by physical processes operating on scales between 100 pc and $\sim$1 kpc, such as feedback from massive stars or AGN activity. This interpretation is supported by multiple lines of evidence. We find that composite-like and AGN-like regions exhibit different slopes compared to \HII-like regions. This aligns with the steeper relation reported by \cite{Gao2025} in early-type galaxies—systems predominantly characterized by such ionization conditions—compared to star-forming galaxies. Furthermore, the scale-dependent variation is significantly weaker when the analysis is restricted to the high-$b$ subsample, reinforcing the connection to local ionization physics. Collectively, our findings provide a more granular understanding of how local physical conditions shape the integrated scaling relations observed in galaxies.

In addition to W3, the WISE W4 band, centered at $\sim22\mu{\rm m}$ and thus comparable to JWST F2100W, has also been previously used to trace dust emission in galaxies. One notable result is the smaller intrinsic scatter in the scaling relation between CO(2-1) and the W3 band compared to that between CO(2-1) and the W4 band reported in \cite{Gao2019} and \cite{Whitcomb2023}. 
However, our work finds that all three MIR bands exhibit similar intrinsic scatter in their correlations with $\ICO$. Our tests show that this conclusion is robust regardless of whether the tightness of the correlation is assessed using intrinsic scatter or correlation coefficient, and it is unaffected by the specific methods used for regression and correlation coefficient analysis. In fact, similar correlation coefficients across different MIR bands are also reported for four PHANGS galaxies in \cite{Leroy2023}. As shown in \autoref{fig:vs_resolution_HII} (bottom panels), although $\sigKS$ varies with resolution, no significant differences are detected between the three bands across all resolutions investigated. 
The remaining plausible explanations for the differences between our results and those previously obtained using W3 and W4 involve variations in instruments and/or sample selection. Resolving this issue requires a direct comparison of WISE and JWST data on the same sample, which should be a focus of future work.

\subsubsection{MIR fluxes from JWST}

By applying median orthogonal distance regression (\modr) on four PHANGS galaxies (IC5332, NGC0628, NGC1365, NGC7496), \cite{Leroy2023} analyze the correlation of CO(2-1) with four MIR bands from JWST: $\blueband$, ${\rm F1000W}$, $\greenband$, and $\redband$. They report that, in all four galaxies, $\redband$ exhibits a sublinear slope ($< 1$), which is significantly flatter than those of the other three bands. This result disagrees with our findings, where the differences in slopes between $\blueband$, $\greenband$, and $\redband$ are slight, and all three bands show superlinear slopes ($> 1$), as presented in \autoref{fig:kbs_scatter}. 
As demonstrated in \citetalias{Jing2024}, this discrepancy arises from the limitations of the \modr\ technique, which tends to produce flatter slopes when the MIR bands have limited S/N and is sensitive to the presence of outliers (e.g., deviations from log-log linear correlations in the brightest regions, as discussed in \autoref{sec:non_log_linear}). These issues affect all bands analyzed. Consequently, the slopes reported in \cite{Leroy2023} are systematically flatter than our results and those in \citetalias{Jing2024}. The significant flattening of the $\redband$ slope compared to other bands is due to its lower S/N and stronger deviations from log-log linear correlations in the brightest regions. 
This highlights the importance of carefully selecting an appropriate regression technique when analyzing datasets with limited S/N and outliers.

\begin{figure*}[ht!]
    % polished
    \centering
    \includegraphics[width=0.95\textwidth]{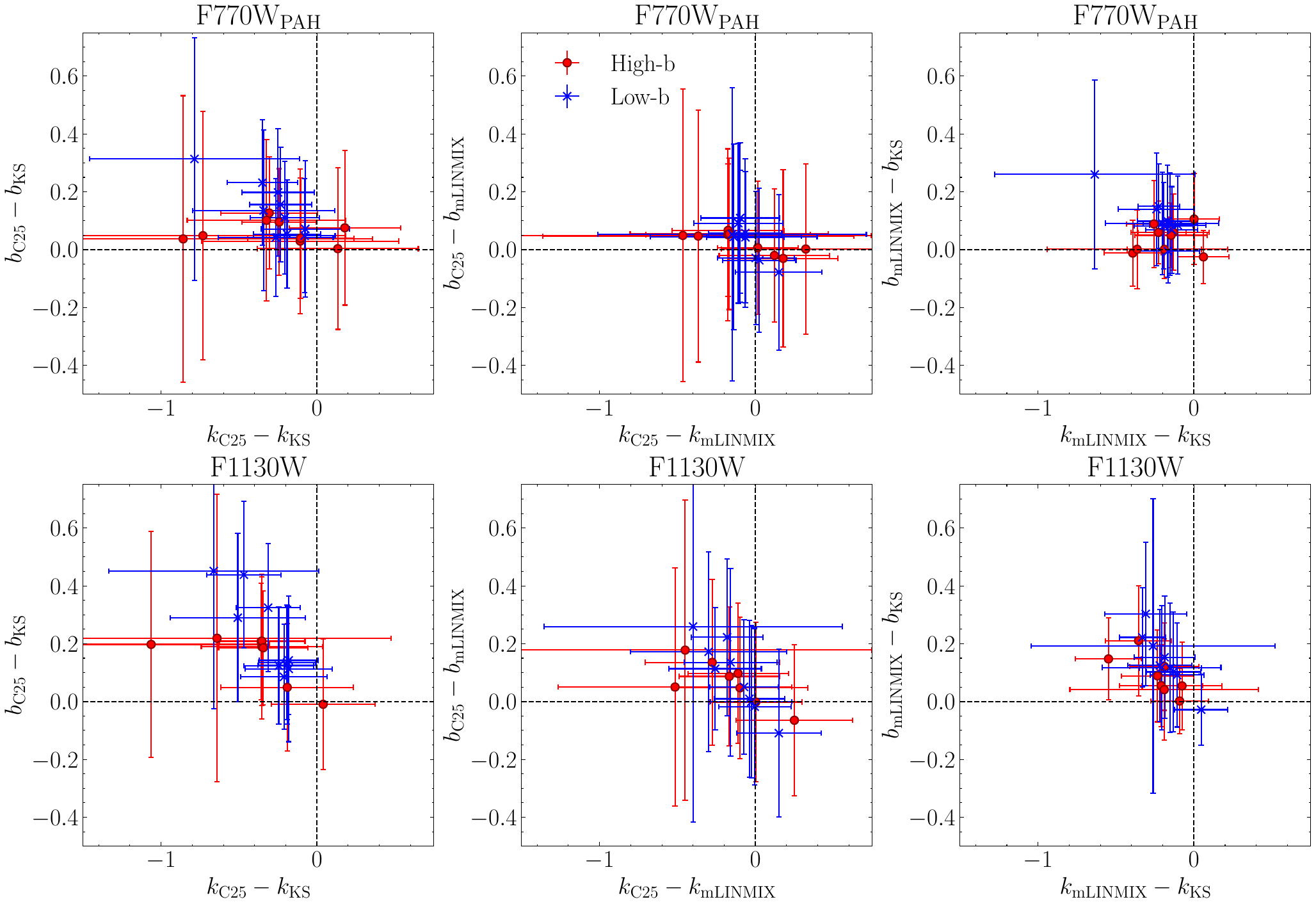}
    \caption{Comparison of the slope ($k$) and intercept ($b$) reported by \cite{Chown2024} ($k_{\rm C24}$, $b_{\rm C24}$) with those derived in this work using \ksmethod\ ($\kKS$, $\bKS$) and \mlinmix\ ($k_{\rm mLINMIX}$, $b_{\rm mLINMIX}$). From left to right, the columns compare \cite{Chown2024} with \ksmethod, \cite{Chown2024} with \mlinmix, and \ksmethod\ with \mlinmix, respectively. The top row shows results for $\blueband$, while the bottom row corresponds to $\greenband$. In each panel, the x-axis represents the differences in $k$, and the y-axis represents the differences in $b$. Galaxies classified as \highb\ are marked as red dots, and those classified as \lowb\ are shown as blue crosses.
    \label{fig:vs_C24}}
\end{figure*}

More recently, \cite{Chown2024} investigate the correlations of $\ICO$ with $\Iblue$ and $\Igreen$ using a much larger sample of 66 galaxies from PHANGS (including the 19 galaxies analyzed in this work). They apply the \linmix\ regression method \citep{Kelly07} on binned data, here referred to as ``\mlinmix''. Although they do not apply line ratio-based classifications, their fitting results for spaxels outside galaxy centers are broadly comparable to those for spaxels with \HII-like ionization conditions in most of our galaxies. In the left column of \autoref{fig:vs_C24}, we compare our best-fit slope $\kKS$ and intercept $\bKS$ with the values ($k_{\text{C25}}$, $b_{\text{C25}}$) reported by \cite{Chown2024}. The best-fit parameters are consistent within $1 - 2 \sigma$, and no distinct behavior emerges between \highb\ and \lowb\ galaxies. However, a systematic underestimation of slope and overestimation of intercept by \cite{Chown2024} is evident. The differences in $k$ and $b$ are anti-correlated, consistent with the well-known $k$-$b$ degeneracy. 

In addition to the different regression methods, differences in sample selection may also contribute to the discrepancies between our results and those in \cite{Chown2024}. The sample differences can be attributed to two key factors: (1) To subtract the continuum in $\Iblueraw$ and calculate the uncertainties of $\Iblue$, we restrict our analysis to spaxels with $\Icont$ coverage. Furthermore, the classification of ionization conditions requires the additional coverage provided by PHANGS-MUSE data. In contrast, \cite{Chown2024} do not utilize PHANGS-MUSE data and estimate the continuum contribution for spaxels without $\Icont$ coverage by using the median ratio of $0.12 \Icont / \Iblueraw$ \footnote{This formula is applied to the 19 galaxies used in our work (Cycle 1). For Cycle 2 galaxies, \cite{Chown2024} estimate the continuum contribution as $0.22 I_{\rm F300W}$.} in spaxels with $\Icont$ coverage, enabling larger spatial coverage. (2) We select spaxels with \HII-like ionization conditions using optical line ratios, whereas \cite{Chown2024} include all spaxels outside galaxy centers. To evaluate the impact of sample differences, we apply \mlinmix\ to our sample and compare the results with those reported by \cite{Chown2024}, as shown in the center column of \autoref{fig:vs_C24}. No systematic differences are found for $\blueband$; however, for $\greenband$, the $k$ ($b$) values reported by \cite{Chown2024} are systematically smaller (larger) than those obtained with \mlinmix\ on our sample. This discrepancy may be related to the radial dependence of the $\ICO$-$\Igreen$ correlation, which warrants further investigation in future work.

The right column of \autoref{fig:vs_C24} directly compares the results obtained using \mlinmix\ and the \ksmethod\ on our sample. As shown, \mlinmix\ tends to yield a flatter slope compared to the  \ksmethod\ in both F770W and F1130W. This agrees with the expectation that binning a noisy x-axis variable (as done by \mlinmix) biases the slope toward zero (e.g., \citealt{Fuller2009}; \citetalias{Jing2024}). Additionally, outliers in the brightest regions, which exhibit lower $\ICO$ than the overall correlation predicts (as discussed in \autoref{sec:non_log_linear}), also leads to underestimated slope. Due to the $k$-$b$ degeneracy, \mlinmix\ also produces larger $b$ values. In summary, while sample differences account for part of the discrepancy between our results and those of \cite{Chown2024}, the remaining differences can be largely attributed to the treatment of x-axis uncertainties and outliers in \mlinmix.

\cite{Chown2024} also report variations in the scaling relationships when spaxels are divided into central and disk regions or into morphologically selected \HII\ regions and diffuse regions. This finding aligns with the ionization condition dependencies identified in our work. However, the specific differences and connections between these dependencies require further investigation in future studies.

Finally, \cite{Chown2024} detect galaxy-to-galaxy variations and identify the intercept $b$ (their $C$) as the dominant contributor to these variations. As discussed in \autoref{sec:res_each_galaxy}, we confirm these results, and further reveal that the correlation between intercept $b$ and global $\sSFR$ likely originates from the observed bimodality in intercept $b$ values.

% As shown before, the best fitting parameters of scaling relation heavily dependent on ionization condition and have strong galaxy-galaxy variation. Thus, the quantitative comparison with previous work should be done with similar sample and ionization condition. 

\subsection{Plausible Physics Implied by the Scaling Relations} \label{sec:physics_behind}

We attempt to understand the underlying physics behind the scaling relations obtained in this work. For this purpose, we express $\ICO$, $\Ired$, $\Igreen$ and $\Iblue$, in terms of more physical quantities:
$$
\begin{aligned}
& \ICO = \betaCO \ngas, \\ 
& \Ired = \betared \frac{\ndust}{\ngas} \ngas, \\
& \Igreen = \betagreen \frac{\nPAH}{\ndust} \frac{\ndust}{\ngas} \ngas, \\
& \Iblue = \betablue \frac{\nPAH}{\ndust} \frac{\ndust}{\ngas} \ngas, 
\end{aligned}
$$
where $\betaCO$, $\betared$, $\betagreen$ and $\betablue$ are the \edit{effective emissivities} (intensity per unit volume density) of $\ICO$, $\Ired$, $\Igreen$ and $\Iblue$, and terms $\ngas$, $\ndust$, and $\nPAH$ represent the density of cold gas, dust, and PAH. To the first order, the correlations of $\ICO$ with $\Iblue$, $\Igreen$, and $\Ired$ arise from their common dependence on $\ngas$. However, the specific values of the slope, intercept, and intrinsic scatter are determined by variations in \edit{effective emissivities}, the dust-to-gas density ratio, and the PAH-to-dust density ratio. Based on this framework, we now discuss the plausible physical mechanisms that may be implied from the key results obtained in this work.

% In this sense, the dependences of correlation on ionization condition and correlation galaxy-galaxy variation imply the variations of emissivities, dust-to-gas ratio, and PAH-to-dust ratio under different ionization conditions and in different galaxies.

\subsubsection{Implications of Ionization Condition Dependence} \label{sec:discuss_ion_cond}

Ionization condition dependencies reflect the influence of the ionization source on the \edit{effective emissivities}, the dust-to-gas ratio, and the PAH-to-dust ratio. The scaling relations observed in \HII-like regions capture these physical properties in typical star-forming environments. \edit{The steeper slope $\kKS$ in AGN-like regions may be associated with the destruction of PAHs and dust grains under strong, hard ionizing radiation fields \citep[e.g.,][]{Draine1979_DustGrain, Tazaki2020_DustAGNa, Tazaki2020_DustAGNb, Zhou2023, Guo2025, Battisti2025, Liu2025, Fischer2025, Katayama2025}. However, the intense radiation fields and warmer dust-heating environments associated with AGN activity may also enhance effective emissivities of PAHs and dust, making the net impact of AGN activity on the MIR emission more complex. The relative contributions of these competing effects to the observed CO--MIR scaling relations require further investigation.} In composite regions, in addition to PAH and dust destruction, variations in $\betaCO$ may also play a significant role. Evidence for $\betaCO$ variation comes from the observed changes in $\alpha_{\rm CO}$ and the $I_{\rm CO(2-1)} / I_{\rm CO(1-0)}$ ratio across different galaxies and within various substructures of individual galaxies \citep[e.g.,][]{Maeda2022, Luo2025, Keenan2025, Komugi2025, Koda2025, Sun2025}.

\subsubsection{Implications of the $\bKS$ Bimodality} \label{sec:discuss_gal_gal_var}

As we showcase in \autoref{sec:res_each_galaxy}, the galaxy-galaxy variation of the scaling relation is primarily led by $\bKS$ and secondly contributed by $\kKS$ and $\sigKS$. The most obvious feature of $\bKS$ is bimodality, which is related to overall host galaxy star formation strength: the \lowb\ (\highb) galaxies tend to have higher (lower) sSFR and star formation efficiency (SFE). The intenser star formation in \lowb\ galaxies would enhance UV radiation filed, which  would heat dust to a warmer temperature and thus enhance $\betared$. The main emission mechanism of PAH is the immediate reradiation of absorbed UV photons. The stronger UV filed can enhance $\betablue$ and $\betagreen$ naturally. 
It is important not to confuse the ``stronger UV radiation field'' discussed here with the ``strong, hard ionizing radiation field'' associated with AGN-like regions. The former refers to a relatively stronger field that enhances PAH and dust emission without significantly destroying them. In contrast, the latter describes an extremely intense radiation field with a hard spectrum, capable of effectively destroying both dust and PAHs. This distinction highlights the complex effects of radiation fields on CO, PAH, dust, and their coupling. These enhancements are manifested primarily in $\bKS$ and secondly in $\kKS$. The higher $\sigKS$ in \lowb\ galaxies is plausibly related to more busty star formation in galaxies with stronger star formation. 

\edit{Given the lack of any significant correlation between the bimodality and SFR, despite SFR being closely linked to UV production, this interpretation may seem counterintuitive at first glance. However, it is important to recognize that the SFR, sSFR, and SFE discussed here are measured on galaxy-wide scales, whereas PAHs and dust respond to the local UV radiation field and physical conditions in their surrounding environments \citep{Draine2021}. Therefore, galaxy-wide sSFR and SFE, which normalize the SFR by stellar mass and molecular gas mass, respectively, may better reflect the UV radiation field experienced by PAHs and dust than the total SFR alone.}

In terms of details, this mechanism can be divided into two scenarios. The first one is that in the galaxies with stronger star formation, there are more \HII\ regions and consequently more leaky photons from them. Therefore, these galaxies have stronger background UV filed, which enhance dust and PAH emission in the whole galaxy. Another scenario is that the \HII\ regions in the galaxies with stronger star formation have higher star formation efficiency. The dust and PAH emission are mainly enhanced in these regions. In reality, these two scenarios cloud coexist, while the comparison between scaling relations obtained in different scale suggest the stronger background UV filed case is more dominated (see \autoref{sec:diff_spatial_scale} for details).

We would like to emphasize that these are reasonable inferences based on the current results, but they have not been confirmed by conclusive evidence. Moreover, it is also rather difficult to explain why it is a $\bKS$ bimodality rather than a continuum.  Therefore, more works should be done in future to explore the exact origin of scaling relation galaxy-galaxy variation and $\bKS$ bimodality.

\subsubsection{Implications of Band-to-Band Variation} \label{sec:discuss_diff_xbands}

After addressing the limited S/N, particularly in $\Ired$, and mitigating the impact of outliers, we find that although the differences in slopes $\kKS$ of PAH bands ($\blueband$ and $\greenband$) and dust band ($\redband$) are detectable, they are minor compared to the galaxy-to-galaxy variation. This suggests that within a given galaxy and under a specific ionization condition, the variation in $\dfrac{\nPAH}{\ndust}$ exists but is not significant across regions of varying brightness. The different intercepts $\bKS$ reflect the overall difference in the $\ICO/\Iblue$, $\ICO/\Igreen$, and $\ICO/\Ired$ ratios. The similar $\sigKS$ values across different MIR bands found in this study indicate that the coupling strength between CO and PAHs, as well as that between CO and dust (at least the hot dust traced by $\Ired$), is comparable in our sample. 

%It is important to note that comparisons of the tightness across different MIR bands reported in various studies are inconsistent, as discussed in \autoref{sec:compare_to_literatures}. These inconsistencies warrant further investigation, as only with additional studies can reliable conclusions be drawn.

\subsubsection{Implications of the Deviations from The Log–log Linear Relation} \label{sec:discuss_non_log_linear}

As we show in \autoref{sec:non_log_linear}, although a log-log linear formula effectively describes the overall correlation of CO with PAH and dust, significant deviations from this relation are evident in the brightest regions. The different degrees of deviation from the log–log linear relation between the PAH bands ($\blueband$ and $\greenband$) and the dust band ($\redband$) suggest that these deviations primarily originate from variations associated with PAH and dust properties. For the PAH bands, a plausible explanation is the multi-photon effect, which enhances the PAH emissivity normalized by the radiation field intensity ($\betablue/U$ and $\betagreen/U$, where $U$ is the radiation field intensity) under stronger radiation fields \citep[e.g.,][]{Draine2021, Richie2025}. Similarly, for $\redband$, higher dust temperatures in stronger radiation fields likely contribute to the enhancement of $\betared$. Future detailed studies are necessary to confirm or refine these proposed mechanisms.

% In most of the time, such deviations behave as a lower $\ICO$ than the log-log linear relation shape. This deviation can be modelled as a piecewise log-log linear relationship that has a flatter slope in the brightest regions for 75\% of the cases. 

\subsubsection{Implications of Spatial Scale Dependence} \label{sec:discuss_diff_scale}

As we point out in \autoref{sec:diff_spatial_scale}, the scaling relation measured at different resolutions should not be regarded solely as a mimicry of observational effects. The scaling relation observed on larger scales (lower resolution) is a combined result of multiple aspects: the underlying scaling relation on smaller scales (higher resolution), dependence of the underlying scaling relation on ionization conditions, and the physical relationships among regions within the considered scale. Therefore, comparing scaling relations obtained at different scales provides insights into the physical scales of the mechanisms that regulate the coupling between CO, PAH, and dust. 

The overall decrease in $\kKS$ at scales larger than $\sim100$ pc as particularly seen in \lowb\ galaxies, suggests that CO, PAH, and dust primarily couple within compact components rather than diffuse ones. At smaller scales, the measured scaling relation directly reflects that in dense clumps. At larger scales, in contrast, the measured scaling relation is dominated by the number of clumps within the corresponding area, approaching a linear correlation (equivalent to a unit slope in the log-log linear formula) gradually. The stronger tendency observed for $\Ired$ further supports this, as $\Ired$ is more concentrated on smaller scales compared to $\Iblue$ and $\Igreen$, as revealed by power spectrum analysis of the PHANGS sample \citep{Charlie2025}.

\edit{The gradual weakening of the $\bKS$ bimodality from sub-100 pc to larger spatial scales} suggests that the corresponding physical mechanism operates on cloud or sub-cloud scales ($\leq 100$ pc). This supports the ``stronger background UV field'' scenario as the origin of $\bKS$ bimodality, rather than the ``\HII\ regions with higher star formation efficiency'' scenario (see \autoref{sec:discuss_gal_gal_var} above for an introduction of these two scenarios). While the background UV field exists throughout a galaxy, it is negligible in the brightest areas of \HII\ regions (typically at their centers). Instead, it significantly enhances dust and PAH emissions in relatively faint areas of \HII\ regions (usually at their outskirts). On scales smaller than 100 pc, the scaling relations resolve both the bright and faint regions, making the $\bKS$ bimodality clearly visible. On larger scales, the measured emissions for each resolution element are dominated by the brightest regions, rendering the $\bKS$ bimodality undetectable. If the ``higher star formation efficiency \HII\ regions'' scenario were the dominant mechanism, the $\bKS$ bimodality would remain detectable across different scales.

Beyond the cancellation of random scatter, the intrinsic scatter $\sigKS$ measured at different scales also encodes the physical correlation between nearby regions. However, reliably decoding this information requires careful removal of the effects of random scatter cancellation, which is beyond the scope of this work. We leave this analysis to future studies.

\subsubsection{Prospects for Future Studies} \label{sec:discuss_future_use}

As discussed above, the statistical results obtained in this work qualitatively relate to a wide range of physical processes. However, directly deriving \edit{effective emissivities}, the dust-to-gas ratio, and the PAH-to-dust ratio remains challenging due to degeneracies among these properties. To fully realize their potential, we propose comparing the statistical results with simulations that incorporate detailed molecular gas, dust, and PAH physics \citep[e.g.,][]{Narayanan2023} to constrain the underlying physical processes. The statistical results presented in this work, with careful treatment of noise and outliers, also facilitate a more direct and fair comparison with such simulations.

\section{Summary} \label{sec:summary}

By applying the recently developed regression technique, \raddest, which effectively handles both uncertainties and outliers in observational data while achieving accurate parameter estimation, we analyze the correlations of CO ($\ICO$) with PAH ($\Iblue$ and $\Igreen$) and dust ($\Ired$) emissions in 19 PHANGS galaxies at spatial scales down to $\lesssim$ 100 pc. We adopt a log-log linear formula to model these scaling relations, and examine the dependence of the best-fit model parameters (slope, intercept and intrinsic scatter) on ionization conditions, host galaxy properties, and spatial scales. Our key findings are summarized as follows:

\begin{itemize}
    \item We confirm that log-log linear scaling relations of $\ICO$ with $\Iblue$, $\Igreen$, and $\Ired$ persist at the 100 pc scale (\autoref{fig:three_cases}, left column of \autoref{fig:all_vs_each}). However, deviations from log-log linearity are observed in regions with the brightest dust or PAH emissions (\autoref{fig:three_cases}).

    \item The scaling relations exhibit a strong dependence on the ionization conditions (\HII-like, composite-like, and AGN-like; see the second-to-last columns of \autoref{fig:all_vs_each}). This highlights the critical role of ionization conditions in regulating the interplay between CO, PAH, and dust in the ISM (refer to \autoref{sec:discuss_ion_cond} for discussion).

    \item Even within the same ionization conditions, significant galaxy-to-galaxy variations in the scaling relations persist (\autoref{fig:kbs_scatter}). These variations are primarily driven by differences in the intercept $b$, which exhibits a bimodal distribution (\autoref{fig:k_vs_b}). This bimodality correlates with the overall star formation strength of the host galaxy (\autoref{fig:b_bimodality}). A plausible explanation is that galaxies with stronger star formation activity have enhanced UV radiation background field, which increase PAH and dust effective emissivities (see \autoref{sec:discuss_gal_gal_var} for further discussion).

    \item No significant ($3\sigma$) correlations are detected between global galaxy properties ($\log \SFR$, $\log \Mstar$, $\log \sSFR$, $\log \LCO$, and $\log \MHI$) and $\kKS$, $\bKS$, or $\sigKS$ (\autoref{fig:vs_property_HII}). This lack of correlation could result from the limited sample size but also suggests that global galaxy properties are not the dominant drivers of the observed galaxy-to-galaxy variations. Although previously reported anti-correlations between $\bKS$ and $\log \sSFR$ are confirmed in $2\sigma$ sense, our results indicate that this trend likely arises from the bimodality of $\bKS$ rather than a continuous correlation (\autoref{fig:vs_sSFR_HII}).

    \item When comparing across different MIR bands ($\Iblue$, $\Igreen$, and $\Ired$), the variations in median $\kKS$ and $\sigKS$ are detectable but minor relative to the galaxy-to-galaxy differences (\autoref{fig:kbs_scatter}). In contrast, $\bKS$ exhibits significant differences across bands. The potential underlying physical mechanisms are discussed in \autoref{sec:discuss_diff_xbands}.

    \item For \HII-like regions, deviations from log-log linearity are well characterized by a flattening of the slope in the brightest regions in 75\% of cases. The degrees of deviations from the log–log linear relation of $\blueband$ and $\greenband$ is similar, whereas $\redband$ exhibits stronger deviations (\autoref{fig:non_loglinear_scatter}, \autoref{fig:dk_vs_q0}). No significant ($3\sigma$) correlations are found between degrees of deviation from the log–log linear relation and global galaxy properties (\autoref{fig:nonlinear_vs_property_HII}). The slope flattening in the brightest regions may result from enhanced PAH and dust \edit{effective emissivities} normalized by the radiation field intensity in regions with stronger radiation fields (see \autoref{sec:discuss_non_log_linear} for discussion).

    \item The parameters $\kKS$, $\bKS$, and $\sigKS$ all depend on the spatial scale of measurement, indicating that the coupling between CO, PAH, and dust are regulated by different mechanisms at varying spatial scales (\autoref{fig:vs_resolution_HII}, see \autoref{sec:discuss_diff_scale} for discussion).
\end{itemize}

% It implies the difference of overall $\ICO/\Iblue$, $\ICO/\Igreen$, and $\ICO/\Ired$ ratios are the dominant of the difference of scaling relations across three x-axis bands. 

The comparison to previous literature is discussed in \autoref{sec:compare_to_literatures}. Leveraging the high-quality dataset provided by the PHANGS project and the improved handling of noise and outliers through the \raddest\ method, this work provides a detailed statistical description of the correlations of CO with PAH and dust. However, several unresolved issues require further investigation: the exact origin of galaxy-to-galaxy variations, particularly the intercept bimodality; the cause of the deviations from the log–log linear relation; and the cause of the inconsistent results regarding the comparison of the $\ICO$-$\Igreen$(or WISE W3 band) and $\ICO$-$\Ired$(or WISE W4 band) scaling relations tightness. 

This analysis is based on a relatively small sample of 19 main sequence galaxies and should be regarded as a description of the behavior within these and similar galaxies. Future studies that extend this analysis to a larger and more representative sample would provide more robust and comprehensive results. As discussed in \autoref{sec:discuss_future_use}, the full potential of these findings can be realized by comparing them with simulations that incorporate detailed molecular gas, dust, and PAH models. Such comparisons would offer valuable constraints and lead to a deeper understanding of the underlying physics driving galaxy evolution. The regression technique \raddest, which demonstrates clear advantages in handling datasets with limited S/N and outliers in this work and \citetalias{Jing2024}, can also be applied to other astrophysical analyses to provide more accurate and unbiased statistical foundations.

%% Please use the acknowledgment and contribution environments. This will 
%% be anonomyized when the "anonymous" style option is used. 
\begin{acknowledgments}

\edit{We are grateful to the anonymous referee and AAS editors for their comments, which have helped us to improve this paper.} 
TJ thanks Stijn Wuyts, Amélie Saintonge, Di Li, Yang Gao, and Fuheng Liang for helpful discussions. This work is supported by the National Key R\&D Program of China (grant NO. 2022YFA1602902), the National Natural Science Foundation of China (grant Nos. 12433003, 11821303, 11973030), and the China Manned Space Program with grant no. CMS-CSST-2025-A10.

This work is based on observations taken as part of the PHANGS large program, including PHANGS-ALMA, PHANGS-JWST, and PHANGS-MUSE. 

This paper makes use of the ALMA data, publically available at ADS/JAO.ALMA (\#2012.1.00650.S, \#2013.1.00803.S, \#2013.1.01161.S, \#2015.1.00121.S, \#2015.1.00782.S, \#2015.1.00925.S, \#2015.1.00956.S, \#2016.1.00386.S, \#2017.1.00392.S, \#2017.1.00766.S, \#2017.1.00886.L, \#2018.1.00484.S, \#2018.1.01321.S, \#2018.1.01651.S, \#2018.A.00062.S, \#2019.1.01235.S, \#2019.2.00129.S). ALMA is a partnership of ESO (representing its member states), NSF (USA) and NINS (Japan), together with NRC (Canada), MOST and ASIAA (Taiwan), and KASI (Republic of Korea), in cooperation with the Republic of Chile. The Joint ALMA Observatory is operated by ESO, AUI/NRAO and NAOJ. The National Radio Astronomy Observatory is a facility of the National Science Foundation operated under cooperative agreement by Associated Universities, Inc.

This work is based in part on observations made with the NASA/ESA/CSA James Webb Space Telescope. The data were obtained from the Mikulski Archive for Space Telescopes at the Space Telescope Science Institute, which is operated by the Association of Universities for Research in Astronomy, Inc., under NASA contract NAS 5-03127 for JWST. These observations are associated with program 2107.

This work is based on data products created from observations collected at the European Organisation for Astronomical Research in the Southern Hemisphere under ESO programme(s) 1100.B-0651, 095.C-0473, and 094.C-0623 (PHANGS-MUSE; PI Schinnerer), as well as 094.B-0321 (MAGNUM; PI Marconi), 099.B-0242, 0100.B-0116, 098.B-0551 (MAD; PI Carollo) and 097.B-0640 (TIMER; PI Gadotti). This research has made use of the services of the ESO Science Archive Facility.

The authors acknowledge the Tsinghua Astrophysics High-Performance Computing platform at Tsinghua University for providing computational and data storage resources that have contributed to the research results reported within this paper.
\end{acknowledgments}

\begin{contribution}
%%This section gives authors the space to recognize author contributions. The text inside this environment is NOT counted towards the total word quanta. At a minimum, manuscripts are expected to include this text:

TJ performed all the data analysis, made all the figures and carried out the first version of the manuscript. CL supervised the whole project, and thoroughly edited the manuscript.

%% But authors are expected to provide more specific details, e.g. 
%%
%%SC was responsible for writing and submitting the manuscript.
%%WWM came up with the initial research concept and edited the manuscript.
%%OTS obtained the funding and edited the manuscript.
%%EBF provided the formal analysis and validation. He also edited the manuscript.
%%GEH Supervised the undergraduates, wrote the software and administers the project github and Zenodo repositories.
%%
%% Authors can use the Contributor Role Taxonomy (CRediT) at
%% https://credit.niso.org
%% for ideas on how write a good statement tailored to their needs.

\end{contribution}

%% To help institutions obtain information on the effectiveness of their 
%% telescopes the AAS Journals has created a group of keywords for telescope 
%% facilities.
%
%% Following the acknowledgments section, use the following syntax and the
%% \facility{} or \facilities{} macros to list the keywords of facilities used 
%% in the research for the paper.  Each keyword is check against the master 
%% list during copy editing.  Individual instruments can be provided in 
%% parentheses, after the keyword, but they are not verified.
\facilities{JWST(MIRI, NIRCam), VLT(MUSE), ALMA}

%% Similar to \facility{}, there is the optional \software command to allow 
%% authors a place to specify which programs were used during the creation of 
%% the manuscript. Authors should list each code and include either a
%% citation or url to the code inside ()s when available.
\software{Python3\citep{python3}, NumPy\citep{numpy}, Matplotlib\citep{matplotlib}, Astropy\citep{astropy:2013, astropy:2018, astropy:2022}, SciPy\citep{scipy}, reproject \citep{reproject_0p13p1}, raddest \citep{Jing2024}, LINMIX \citep{Kelly07}}

%% Appendix material should be preceded with a single \appendix command.
%% There should be a \section command for each appendix. Mark appendix
%% subsections with the same markup you use in the main body of the paper.
%%
%% Each Appendix (indicated with \section) will be lettered A, B, C, etc.
%% The equation counter will reset when it encounters the \appendix
%% command and will number appendix equations (A1), (A2), etc. The
%% Figure and Table counter will not reset.

\bibliography{sample7}{}
\bibliographystyle{aasjournalv7}

\appendix

\section{Results of Single Log-log Linear Regression Analysis} \label{app:res}

\edit{Best-fit results based on \ksmethod\ for different MIR bands in regions with varying ionization conditions, derived from the combined sample of all 19 galaxies, are presented in \autoref{tab:fit_res_all}. Corresponding results obtained separately for each galaxy are listed in \autoref{tab:fit_res}.}

\begin{deluxetable}{c@{\hspace{1pt}}ccccccccc}
	\tablecolumns{10}
	% \tablewidth{\textwidth}
    \tablewidth{0pt}
	\tablecaption{Best-fit results based on \ksmethod\ for different MIR bands in regions with varying ionization conditions, derived from the combined sample of all 19 galaxies \label{tab:fit_res_all}}
	\tablehead{
        \colhead{Ionization} &
		% \colhead{\multirow{2}{*}{Ionization Condition}} &
		\multicolumn{3}{c}{$\blueband$} &
        \multicolumn{3}{c}{$\greenband$} &
        \multicolumn{3}{c}{$\redband$}
        \\
		\colhead{Condition} &
		\colhead{$\kKS$} &
        \colhead{$\bKS$} &
		\colhead{$\sigKS$} &
        \colhead{$\kKS$} &
        \colhead{$\bKS$} &
		\colhead{$\sigKS$} &
        \colhead{$\kKS$} &
        \colhead{$\bKS$} &
		\colhead{$\sigKS$}
	}
	\startdata
    all & $1.12 \pm 0.11$ & $-0.04 \pm 0.05$ & $0.37 \pm 0.09$ & $1.17 \pm 0.12$ & $-0.30 \pm 0.06$ & $0.35 \pm 0.09$ & $1.10 \pm 0.11$ & $0.12 \pm 0.04$ & $0.29 \pm 0.11$ \\
    H{\sc ii}-like & $1.12 \pm 0.11$ & $-0.11 \pm 0.05$ & $0.31 \pm 0.09$ & $1.19 \pm 0.10$ & $-0.35 \pm 0.06$ & $0.31 \pm 0.09$ & $1.06 \pm 0.12$ & $0.05 \pm 0.04$ & $0.26 \pm 0.09$ \\
    composite-like & $1.27 \pm 0.17$ & $0.08 \pm 0.06$ & $0.39 \pm 0.11$ & $1.26 \pm 0.17$ & $-0.27 \pm 0.06$ & $0.38 \pm 0.11$ & $1.53 \pm 0.14$ & $0.24 \pm 0.05$ & $0.21 \pm 0.13$ \\
    AGN-like & $2.17 \pm 0.21$ & $0.31 \pm 0.09$ & $0.69 \pm 0.13$ & $1.84 \pm 0.21$ & $-0.33 \pm 0.09$ & $0.60 \pm 0.15$ & $1.78 \pm 0.20$ & $0.27 \pm 0.09$ & $0.50 \pm 0.18$ \\
	\enddata
    % \tablecomments{}  
\end{deluxetable}

\begin{longrotatetable}
\begin{deluxetable*}{l@{\hspace{0pt}}l@{\hspace{1pt}}c@{\hspace{1pt}}ccccccccc}
	\tablecolumns{12}
	\tablewidth{\textwidth}
	\tablecaption{Best-fit results based on \ksmethod\ for different MIR bands in regions with varying ionization conditions, obtained separately for each galaxy \label{tab:fit_res}}
	\tablehead{
        \colhead{\multirow{2}{*}{ID \tablenotemark{a}}} &
		\colhead{\multirow{2}{*}{Name}} &
        \colhead{Ionization} &
		% \colhead{\multirow{2}{*}{Ionization Condition}} &
		\multicolumn{3}{c}{$\blueband$} &
        \multicolumn{3}{c}{$\greenband$} &
        \multicolumn{3}{c}{$\redband$}
        \\
        \colhead{} &
		\colhead{} &
		\colhead{Condition} &
		\colhead{$\kKS$} &
        \colhead{$\bKS$} &
		\colhead{$\sigKS$} &
        \colhead{$\kKS$} &
        \colhead{$\bKS$} &
		\colhead{$\sigKS$} &
        \colhead{$\kKS$} &
        \colhead{$\bKS$} &
		\colhead{$\sigKS$}
	}
	\startdata
    \multirow{4}{*}{1} & \multirow{4}{*}{NGC5068} & all & -- & -- & -- & -- & -- & -- & -- & -- & -- \\
    ~ & ~ & H{\sc ii}-like & $1.70 \pm 0.25$ & $-0.42 \pm 0.11$ & $0.41 \pm 0.23$ & $1.60 \pm 0.27$ & $-0.70 \pm 0.12$ & $0.33 \pm 0.20$ & $1.50 \pm 0.29$ & $-0.20 \pm 0.11$ & $0.39 \pm 0.22$ \\
    ~ & ~ & composite-like & -- & -- & -- & -- & -- & -- & -- & -- & -- \\
    ~ & ~ & AGN-like & $1.82 \pm 0.68$ & $-0.25 \pm 0.29$ & $0.34 \pm 0.22$ & $2.13 \pm 0.79$ & $-0.67 \pm 0.19$ & $0.28 \pm 0.18$ & $1.91 \pm 0.69$ & $0.03 \pm 0.41$ & $0.36 \pm 0.24$ \\
    \hline
    \multirow{4}{*}{2} & \multirow{4}{*}{IC5332} & all & $1.21 \pm 0.31$ & $0.19 \pm 0.10$ & $0.21 \pm 0.14$ & $1.28 \pm 0.31$ & $-0.03 \pm 0.07$ & $0.21 \pm 0.13$ & $1.36 \pm 0.33$ & $0.24 \pm 0.12$ & $0.19 \pm 0.13$ \\
    ~ & ~ & H{\sc ii}-like & $1.15 \pm 0.30$ & $0.15 \pm 0.08$ & $0.19 \pm 0.13$ & $1.10 \pm 0.30$ & $-0.05 \pm 0.06$ & $0.17 \pm 0.12$ & $1.26 \pm 0.28$ & $0.20 \pm 0.09$ & $0.20 \pm 0.13$ \\
    ~ & ~ & composite-like & $1.15 \pm 0.43$ & $0.17 \pm 0.18$ & $0.24 \pm 0.15$ & -- & -- & -- & $1.37 \pm 0.49$ & $0.24 \pm 0.21$ & $0.21 \pm 0.15$ \\
    ~ & ~ & AGN-like & $1.60 \pm 0.61$ & $0.07 \pm 0.37$ & $0.76 \pm 0.25$ & $1.72 \pm 0.78$ & $-0.29 \pm 0.31$ & $0.81 \pm 0.26$ & $1.87 \pm 0.73$ & $0.05 \pm 0.40$ & $0.84 \pm 0.24$ \\
    \hline
    \multirow{4}{*}{3} & \multirow{4}{*}{NGC1087} & all & $1.33 \pm 0.11$ & $-0.37 \pm 0.05$ & $0.25 \pm 0.10$ & $1.37 \pm 0.12$ & $-0.61 \pm 0.06$ & $0.26 \pm 0.10$ & $1.21 \pm 0.10$ & $-0.16 \pm 0.03$ & $0.28 \pm 0.09$ \\
    ~ & ~ & H{\sc ii}-like & $1.37 \pm 0.10$ & $-0.38 \pm 0.05$ & $0.25 \pm 0.08$ & $1.46 \pm 0.11$ & $-0.67 \pm 0.07$ & $0.25 \pm 0.08$ & $1.21 \pm 0.10$ & $-0.17 \pm 0.03$ & $0.29 \pm 0.09$ \\
    ~ & ~ & composite-like & $1.50 \pm 0.31$ & $-0.18 \pm 0.10$ & $0.20 \pm 0.13$ & $1.46 \pm 0.32$ & $-0.50 \pm 0.07$ & $0.20 \pm 0.13$ & $1.71 \pm 0.38$ & $0.10 \pm 0.17$ & $0.27 \pm 0.16$ \\
    ~ & ~ & AGN-like & $-0.37 \pm 1.25$ & $-1.13 \pm 0.71$ & $0.25 \pm 0.17$ & $0.33 \pm 1.26$ & $-0.81 \pm 0.48$ & $0.27 \pm 0.17$ & $-0.98 \pm 2.44$ & $-1.74 \pm 1.71$ & $0.38 \pm 0.25$ \\
    \hline
    \multirow{4}{*}{4} & \multirow{4}{*}{NGC1385} & all & $1.23 \pm 0.09$ & $-0.40 \pm 0.06$ & $0.25 \pm 0.10$ & $1.27 \pm 0.09$ & $-0.64 \pm 0.07$ & $0.26 \pm 0.09$ & $1.00 \pm 0.09$ & $-0.17 \pm 0.04$ & $0.28 \pm 0.09$ \\
    ~ & ~ & H{\sc ii}-like & $1.11 \pm 0.09$ & $-0.37 \pm 0.06$ & $0.27 \pm 0.08$ & $1.25 \pm 0.09$ & $-0.63 \pm 0.07$ & $0.24 \pm 0.09$ & $1.00 \pm 0.09$ & $-0.17 \pm 0.05$ & $0.28 \pm 0.09$ \\
    ~ & ~ & composite-like & $2.16 \pm 0.42$ & $-0.36 \pm 0.09$ & $0.24 \pm 0.15$ & $1.99 \pm 0.30$ & $-0.79 \pm 0.09$ & $0.19 \pm 0.13$ & $2.02 \pm 0.48$ & $-0.16 \pm 0.15$ & $0.50 \pm 0.26$ \\
    ~ & ~ & AGN-like & $0.05 \pm 1.50$ & $-1.36 \pm 0.58$ & $0.39 \pm 0.26$ & $-0.46 \pm 1.86$ & $-1.35 \pm 0.35$ & $0.33 \pm 0.22$ & $1.29 \pm 1.39$ & $-0.72 \pm 0.75$ & $0.40 \pm 0.26$ \\
    \hline
    \multirow{4}{*}{5} & \multirow{4}{*}{NGC2835} & all & $1.21 \pm 0.17$ & $-0.16 \pm 0.05$ & $0.20 \pm 0.11$ & $1.28 \pm 0.18$ & $-0.43 \pm 0.07$ & $0.20 \pm 0.12$ & $1.18 \pm 0.18$ & $-0.04 \pm 0.05$ & $0.19 \pm 0.12$ \\
    ~ & ~ & H{\sc ii}-like & $1.28 \pm 0.16$ & $-0.24 \pm 0.05$ & $0.30 \pm 0.11$ & $1.40 \pm 0.20$ & $-0.50 \pm 0.08$ & $0.25 \pm 0.12$ & $1.17 \pm 0.18$ & $-0.10 \pm 0.05$ & $0.30 \pm 0.11$ \\
    ~ & ~ & composite-like & $1.36 \pm 0.25$ & $0.01 \pm 0.07$ & $0.20 \pm 0.13$ & $1.56 \pm 0.30$ & $-0.34 \pm 0.06$ & $0.17 \pm 0.11$ & $1.89 \pm 0.31$ & $0.38 \pm 0.10$ & $0.14 \pm 0.10$ \\
    ~ & ~ & AGN-like & -- & -- & -- & -- & -- & -- & -- & -- & -- \\
    \hline
    \multirow{4}{*}{6} & \multirow{4}{*}{NGC7496} & all & $1.33 \pm 0.11$ & $-0.05 \pm 0.03$ & $0.33 \pm 0.07$ & $1.38 \pm 0.13$ & $-0.35 \pm 0.04$ & $0.31 \pm 0.07$ & $1.12 \pm 0.09$ & $0.01 \pm 0.03$ & $0.34 \pm 0.07$ \\
    ~ & ~ & H{\sc ii}-like & $1.25 \pm 0.11$ & $-0.11 \pm 0.03$ & $0.28 \pm 0.07$ & $1.26 \pm 0.10$ & $-0.35 \pm 0.04$ & $0.28 \pm 0.06$ & $1.10 \pm 0.10$ & $0.03 \pm 0.03$ & $0.29 \pm 0.06$ \\
    ~ & ~ & composite-like & $1.75 \pm 0.18$ & $0.05 \pm 0.05$ & $0.40 \pm 0.07$ & $1.75 \pm 0.19$ & $-0.38 \pm 0.04$ & $0.38 \pm 0.07$ & -- & -- & -- \\
    ~ & ~ & AGN-like & -- & -- & -- & -- & -- & -- & -- & -- & -- \\
    \hline
    \multirow{4}{*}{7} & \multirow{4}{*}{NGC0628} & all & $0.97 \pm 0.09$ & $0.14 \pm 0.03$ & $0.22 \pm 0.08$ & $1.17 \pm 0.13$ & $-0.15 \pm 0.06$ & $0.21 \pm 0.09$ & $1.16 \pm 0.11$ & $0.29 \pm 0.02$ & $0.15 \pm 0.09$ \\
    ~ & ~ & H{\sc ii}-like & $1.00 \pm 0.10$ & $0.10 \pm 0.03$ & $0.22 \pm 0.07$ & $1.12 \pm 0.11$ & $-0.15 \pm 0.06$ & $0.20 \pm 0.07$ & $1.06 \pm 0.09$ & $0.26 \pm 0.02$ & $0.19 \pm 0.08$ \\
    ~ & ~ & composite-like & $1.20 \pm 0.13$ & $0.21 \pm 0.04$ & $0.20 \pm 0.10$ & $1.19 \pm 0.13$ & $-0.11 \pm 0.05$ & $0.18 \pm 0.09$ & -- & -- & -- \\
    ~ & ~ & AGN-like & $1.75 \pm 1.01$ & $-0.63 \pm 0.47$ & $0.59 \pm 0.36$ & $0.39 \pm 0.69$ & $-0.96 \pm 0.25$ & $0.44 \pm 0.27$ & $0.44 \pm 0.74$ & $-0.74 \pm 0.29$ & $0.41 \pm 0.27$ \\
    \hline
    \multirow{4}{*}{8} & \multirow{4}{*}{NGC3351} & all & $1.17 \pm 0.16$ & $0.22 \pm 0.06$ & $0.21 \pm 0.11$ & $1.44 \pm 0.21$ & $-0.11 \pm 0.05$ & $0.24 \pm 0.11$ & -- & -- & -- \\
    ~ & ~ & H{\sc ii}-like & $1.28 \pm 0.15$ & $0.19 \pm 0.03$ & $0.22 \pm 0.08$ & $1.44 \pm 0.18$ & $-0.10 \pm 0.04$ & $0.23 \pm 0.07$ & $1.49 \pm 0.20$ & $0.30 \pm 0.04$ & $0.17 \pm 0.09$ \\
    ~ & ~ & composite-like & $1.22 \pm 0.20$ & $0.27 \pm 0.09$ & $0.22 \pm 0.13$ & $1.48 \pm 0.28$ & $-0.10 \pm 0.06$ & $0.24 \pm 0.14$ & $1.64 \pm 0.28$ & $0.32 \pm 0.11$ & $0.24 \pm 0.13$ \\
    ~ & ~ & AGN-like & $0.51 \pm 0.84$ & $-0.92 \pm 0.73$ & $0.70 \pm 0.43$ & $0.41 \pm 0.84$ & $-0.84 \pm 0.37$ & $0.46 \pm 0.31$ & $1.28 \pm 1.00$ & $-0.56 \pm 0.48$ & $0.71 \pm 0.43$ \\
    \hline
    \multirow{4}{*}{9} & \multirow{4}{*}{NGC4254} & all & $1.27 \pm 0.09$ & $-0.19 \pm 0.06$ & $0.27 \pm 0.07$ & $1.35 \pm 0.09$ & $-0.46 \pm 0.07$ & $0.27 \pm 0.07$ & $1.16 \pm 0.08$ & $0.05 \pm 0.04$ & $0.28 \pm 0.07$ \\
    ~ & ~ & H{\sc ii}-like & $1.35 \pm 0.09$ & $-0.24 \pm 0.07$ & $0.26 \pm 0.08$ & $1.34 \pm 0.09$ & $-0.47 \pm 0.08$ & $0.26 \pm 0.07$ & $1.17 \pm 0.08$ & $0.03 \pm 0.05$ & $0.27 \pm 0.07$ \\
    ~ & ~ & composite-like & $1.25 \pm 0.12$ & $-0.06 \pm 0.04$ & $0.29 \pm 0.08$ & $1.30 \pm 0.12$ & $-0.37 \pm 0.05$ & $0.29 \pm 0.07$ & $1.27 \pm 0.11$ & $0.10 \pm 0.03$ & $0.29 \pm 0.08$ \\
    ~ & ~ & AGN-like & -- & -- & -- & -- & -- & -- & -- & -- & -- \\
    \hline
    \multirow{4}{*}{10} & \multirow{4}{*}{NGC4303} & all & $1.25 \pm 0.10$ & $-0.21 \pm 0.06$ & $0.31 \pm 0.07$ & $1.36 \pm 0.10$ & $-0.56 \pm 0.08$ & $0.30 \pm 0.06$ & $1.14 \pm 0.09$ & $-0.03 \pm 0.05$ & $0.33 \pm 0.06$ \\
    ~ & ~ & H{\sc ii}-like & $1.21 \pm 0.10$ & $-0.27 \pm 0.08$ & $0.28 \pm 0.05$ & $1.30 \pm 0.10$ & $-0.56 \pm 0.10$ & $0.28 \pm 0.05$ & $1.07 \pm 0.09$ & $-0.03 \pm 0.06$ & $0.29 \pm 0.06$ \\
    ~ & ~ & composite-like & $1.65 \pm 0.14$ & $-0.22 \pm 0.05$ & $0.30 \pm 0.08$ & $1.66 \pm 0.15$ & $-0.65 \pm 0.09$ & $0.30 \pm 0.08$ & $1.46 \pm 0.13$ & $-0.01 \pm 0.04$ & $0.31 \pm 0.08$ \\
    ~ & ~ & AGN-like & -- & -- & -- & -- & -- & -- & -- & -- & -- \\
    \hline
    \multirow{4}{*}{11} & \multirow{4}{*}{NGC4535} & all & $1.21 \pm 0.11$ & $0.18 \pm 0.03$ & $0.29 \pm 0.07$ & $1.31 \pm 0.12$ & $-0.08 \pm 0.03$ & $0.29 \pm 0.07$ & $1.29 \pm 0.11$ & $0.25 \pm 0.04$ & $0.28 \pm 0.09$ \\
    ~ & ~ & H{\sc ii}-like & $1.22 \pm 0.10$ & $0.07 \pm 0.03$ & $0.26 \pm 0.06$ & $1.37 \pm 0.13$ & $-0.18 \pm 0.05$ & $0.25 \pm 0.06$ & $1.17 \pm 0.10$ & $0.17 \pm 0.03$ & $0.27 \pm 0.06$ \\
    ~ & ~ & composite-like & $1.73 \pm 0.23$ & $0.37 \pm 0.07$ & $0.28 \pm 0.08$ & $1.68 \pm 0.21$ & $-0.01 \pm 0.03$ & $0.28 \pm 0.09$ & $1.90 \pm 0.20$ & $0.46 \pm 0.07$ & $0.27 \pm 0.09$ \\
    ~ & ~ & AGN-like & -- & -- & -- & -- & -- & -- & -- & -- & -- \\
    \hline
    \multirow{4}{*}{12} & \multirow{4}{*}{NGC1300} & all & $1.09 \pm 0.16$ & $0.09 \pm 0.06$ & $0.33 \pm 0.10$ & $1.33 \pm 0.19$ & $-0.15 \pm 0.05$ & $0.32 \pm 0.11$ & -- & -- & -- \\
    ~ & ~ & H{\sc ii}-like & $1.22 \pm 0.13$ & $0.01 \pm 0.03$ & $0.25 \pm 0.08$ & $1.23 \pm 0.12$ & $-0.19 \pm 0.04$ & $0.24 \pm 0.08$ & $1.13 \pm 0.11$ & $0.15 \pm 0.03$ & $0.21 \pm 0.09$ \\
    ~ & ~ & composite-like & $1.82 \pm 0.30$ & $0.41 \pm 0.15$ & $0.45 \pm 0.14$ & $1.70 \pm 0.27$ & $-0.01 \pm 0.09$ & $0.31 \pm 0.17$ & -- & -- & -- \\
    ~ & ~ & AGN-like & $2.10 \pm 0.51$ & $0.67 \pm 0.38$ & $0.56 \pm 0.29$ & $1.71 \pm 0.36$ & $0.06 \pm 0.21$ & $0.48 \pm 0.27$ & $1.86 \pm 0.47$ & $0.38 \pm 0.33$ & $0.58 \pm 0.29$ \\
    \hline
    \multirow{4}{*}{13} & \multirow{4}{*}{NGC1512} & all & $0.85 \pm 0.12$ & $0.11 \pm 0.06$ & $0.18 \pm 0.10$ & $0.98 \pm 0.14$ & $-0.08 \pm 0.05$ & $0.16 \pm 0.10$ & $1.07 \pm 0.15$ & $0.23 \pm 0.08$ & $0.17 \pm 0.10$ \\
    ~ & ~ & H{\sc ii}-like & $1.01 \pm 0.11$ & $0.07 \pm 0.04$ & $0.15 \pm 0.09$ & $1.10 \pm 0.11$ & $-0.11 \pm 0.03$ & $0.14 \pm 0.08$ & $1.15 \pm 0.12$ & $0.22 \pm 0.05$ & $0.13 \pm 0.08$ \\
    ~ & ~ & composite-like & $0.98 \pm 0.15$ & $0.25 \pm 0.09$ & $0.16 \pm 0.10$ & $0.97 \pm 0.17$ & $-0.04 \pm 0.07$ & $0.15 \pm 0.09$ & $1.37 \pm 0.21$ & $0.43 \pm 0.12$ & $0.13 \pm 0.09$ \\
    ~ & ~ & AGN-like & $0.99 \pm 0.32$ & $0.30 \pm 0.28$ & $0.29 \pm 0.20$ & $0.63 \pm 0.23$ & $-0.18 \pm 0.16$ & $0.30 \pm 0.18$ & $1.84 \pm 0.51$ & $0.82 \pm 0.39$ & $0.31 \pm 0.18$ \\
    \hline
    \multirow{4}{*}{14} & \multirow{4}{*}{NGC1672} & all & $1.46 \pm 0.14$ & $-0.25 \pm 0.06$ & $0.39 \pm 0.09$ & $1.63 \pm 0.16$ & $-0.62 \pm 0.08$ & $0.36 \pm 0.09$ & $1.39 \pm 0.13$ & $-0.05 \pm 0.05$ & $0.40 \pm 0.10$ \\
    ~ & ~ & H{\sc ii}-like & $1.31 \pm 0.10$ & $-0.24 \pm 0.05$ & $0.30 \pm 0.08$ & $1.46 \pm 0.12$ & $-0.54 \pm 0.07$ & $0.27 \pm 0.08$ & $1.17 \pm 0.10$ & $-0.04 \pm 0.04$ & $0.31 \pm 0.08$ \\
    ~ & ~ & composite-like & $1.76 \pm 0.27$ & $-0.27 \pm 0.06$ & $0.60 \pm 0.10$ & $1.78 \pm 0.25$ & $-0.71 \pm 0.09$ & $0.58 \pm 0.11$ & $1.52 \pm 0.25$ & $-0.10 \pm 0.07$ & $0.60 \pm 0.10$ \\
    ~ & ~ & AGN-like & -- & -- & -- & -- & -- & -- & -- & -- & -- \\
    \hline
    \multirow{4}{*}{15} & \multirow{4}{*}{NGC4321} & all & $1.26 \pm 0.11$ & $0.08 \pm 0.04$ & $0.32 \pm 0.06$ & $1.46 \pm 0.13$ & $-0.29 \pm 0.06$ & $0.31 \pm 0.06$ & $1.31 \pm 0.11$ & $0.15 \pm 0.03$ & $0.31 \pm 0.06$ \\
    ~ & ~ & H{\sc ii}-like & $1.32 \pm 0.10$ & $-0.02 \pm 0.04$ & $0.28 \pm 0.06$ & $1.49 \pm 0.12$ & $-0.36 \pm 0.07$ & $0.27 \pm 0.06$ & $1.22 \pm 0.10$ & $0.13 \pm 0.04$ & $0.28 \pm 0.06$ \\
    ~ & ~ & composite-like & $1.76 \pm 0.17$ & $0.14 \pm 0.03$ & $0.32 \pm 0.07$ & $1.88 \pm 0.19$ & $-0.37 \pm 0.07$ & $0.30 \pm 0.07$ & $1.95 \pm 0.16$ & $0.24 \pm 0.03$ & $0.28 \pm 0.08$ \\
    ~ & ~ & AGN-like & -- & -- & -- & -- & -- & -- & -- & -- & -- \\
    \hline
    \multirow{4}{*}{16} & \multirow{4}{*}{NGC1566} & all & $1.33 \pm 0.12$ & $-0.25 \pm 0.06$ & $0.43 \pm 0.09$ & $1.44 \pm 0.15$ & $-0.57 \pm 0.08$ & $0.41 \pm 0.09$ & $1.36 \pm 0.19$ & $-0.10 \pm 0.05$ & $0.46 \pm 0.09$ \\
    ~ & ~ & H{\sc ii}-like & $1.35 \pm 0.12$ & $-0.28 \pm 0.07$ & $0.36 \pm 0.08$ & $1.49 \pm 0.14$ & $-0.67 \pm 0.10$ & $0.40 \pm 0.07$ & $1.16 \pm 0.11$ & $-0.07 \pm 0.05$ & $0.40 \pm 0.08$ \\
    ~ & ~ & composite-like & $1.64 \pm 0.22$ & $-0.21 \pm 0.05$ & $0.46 \pm 0.10$ & $1.65 \pm 0.23$ & $-0.67 \pm 0.09$ & $0.49 \pm 0.09$ & $1.77 \pm 0.30$ & $-0.09 \pm 0.06$ & $0.55 \pm 0.12$ \\
    ~ & ~ & AGN-like & $2.11 \pm 0.24$ & $-0.13 \pm 0.13$ & $0.33 \pm 0.18$ & $2.40 \pm 0.20$ & $-1.21 \pm 0.19$ & $0.25 \pm 0.16$ & $2.48 \pm 0.18$ & $-0.48 \pm 0.11$ & $0.32 \pm 0.17$ \\
    \hline
    \multirow{4}{*}{17} & \multirow{4}{*}{NGC3627} & all & $1.27 \pm 0.11$ & $-0.15 \pm 0.06$ & $0.35 \pm 0.06$ & $1.42 \pm 0.12$ & $-0.54 \pm 0.09$ & $0.37 \pm 0.07$ & -- & -- & -- \\
    ~ & ~ & H{\sc ii}-like & $1.25 \pm 0.10$ & $-0.18 \pm 0.07$ & $0.31 \pm 0.06$ & $1.35 \pm 0.11$ & $-0.50 \pm 0.09$ & $0.34 \pm 0.05$ & $1.08 \pm 0.09$ & $0.07 \pm 0.05$ & $0.33 \pm 0.05$ \\
    ~ & ~ & composite-like & -- & -- & -- & $1.46 \pm 0.18$ & $-0.55 \pm 0.09$ & $0.41 \pm 0.06$ & $1.47 \pm 0.16$ & $-0.03 \pm 0.04$ & $0.42 \pm 0.07$ \\
    ~ & ~ & AGN-like & -- & -- & -- & -- & -- & -- & -- & -- & -- \\
    \hline
    \multirow{4}{*}{18} & \multirow{4}{*}{NGC1433} & all & $1.30 \pm 0.16$ & $0.16 \pm 0.07$ & $0.27 \pm 0.14$ & $1.33 \pm 0.20$ & $-0.17 \pm 0.05$ & $0.24 \pm 0.13$ & -- & -- & -- \\
    ~ & ~ & H{\sc ii}-like & $1.11 \pm 0.13$ & $0.02 \pm 0.03$ & $0.17 \pm 0.09$ & $1.17 \pm 0.14$ & $-0.20 \pm 0.04$ & $0.18 \pm 0.09$ & $1.07 \pm 0.12$ & $0.11 \pm 0.04$ & $0.17 \pm 0.09$ \\
    ~ & ~ & composite-like & $1.91 \pm 0.29$ & $0.50 \pm 0.13$ & $0.29 \pm 0.16$ & $1.59 \pm 0.31$ & $-0.10 \pm 0.08$ & $0.24 \pm 0.15$ & $2.33 \pm 0.32$ & $0.75 \pm 0.15$ & $0.17 \pm 0.12$ \\
    ~ & ~ & AGN-like & $1.87 \pm 0.50$ & $0.47 \pm 0.31$ & $0.64 \pm 0.30$ & $1.69 \pm 0.51$ & $-0.24 \pm 0.15$ & $0.46 \pm 0.27$ & $2.01 \pm 0.45$ & $0.39 \pm 0.23$ & $0.51 \pm 0.29$ \\
    \hline
    \multirow{4}{*}{19} & \multirow{4}{*}{NGC1365} & all & $1.61 \pm 0.16$ & $0.22 \pm 0.05$ & $0.49 \pm 0.09$ & $1.81 \pm 0.16$ & $-0.25 \pm 0.05$ & $0.44 \pm 0.09$ & -- & -- & -- \\
    ~ & ~ & H{\sc ii}-like & $1.12 \pm 0.13$ & $-0.04 \pm 0.03$ & $0.25 \pm 0.08$ & -- & -- & -- & $1.02 \pm 0.10$ & $0.14 \pm 0.03$ & $0.23 \pm 0.09$ \\
    ~ & ~ & composite-like & $1.73 \pm 0.17$ & $0.30 \pm 0.05$ & $0.44 \pm 0.09$ & $1.88 \pm 0.17$ & $-0.21 \pm 0.05$ & $0.47 \pm 0.09$ & -- & -- & -- \\
    ~ & ~ & AGN-like & $2.10 \pm 0.27$ & $0.59 \pm 0.04$ & $0.47 \pm 0.09$ & $1.84 \pm 0.23$ & $-0.06 \pm 0.07$ & $0.48 \pm 0.08$ & -- & -- & -- \\
    \hline
	\enddata
    \tablenotetext{a}{As defined in \autoref{tab:galaxies}, ordered by galactic stellar mass        $\Mstar$.}  
    \tablecomments{We discard results based on datasets with fewer than 100 data points or those that fail the goodness-of-fit test for NF, as outlined in \autoref{sec:regression}. Such results are indicated as ``--'' in the table.}  
\end{deluxetable*}
\end{longrotatetable}

%% For this sample we use BibTeX plus aasjournalv7.bst to generate the
%% the bibliography. The sample7.bib file was populated from ADS. To
%% get the citations to show in the compiled file do the following:
%%
%% pdflatex sample7.tex
%% bibtext sample7
%% pdflatex sample7.tex
%% pdflatex sample7.tex

%% This command is needed to show the entire author+affiliation list when
%% the collaboration and author truncation commands are used.  It has to
%% go at the end of the manuscript.
%\allauthors

%% Include this line if you are using the \added, \replaced, \deleted
%% commands to see a summary list of all changes at the end of the article.
%\listofchanges

\end{document}